\documentclass[a4paper,11pt]{article}
\pdfoutput=1 

\usepackage{jcappub} 
\usepackage[T1]{fontenc} 
\usepackage{tikz-feynhand}
\usepackage{graphicx}
\usepackage{xcolor}
\usepackage{caption,subcaption}
\usepackage{mathrsfs,mathtools}
\usepackage{physics,amssymb}
\usepackage{bm}
\usepackage{braket}
\usepackage{listings}
\usepackage{cases}
\usepackage{comment}
\usepackage{soul}
\usepackage{cancel}
\usepackage{cases}
\usepackage[utf8]{inputenc}
\usepackage{url}
\usepackage[normalem]{ulem}
\usepackage{xspace}
\usepackage{aas_macros}
\usepackage{acronym}
\usepackage{siunitx}
\usepackage{gensymb}

\hypersetup{colorlinks=true
,urlcolor=DARKBLUE
,anchorcolor=DARKBLUE
,citecolor=DARKBLUE
,filecolor=DARKBLUE
,linkcolor=DARKBLUE
,menucolor=DARKBLUE
,linktocpage=true
,pdfproducer=medialab
,pdfa=true
}

\usepackage{xcolor}
\definecolor{bleudefrance}{rgb}{0.19, 0.55, 0.91}
\definecolor{desyblue}{HTML}{009EE2}
\definecolor{desyorange}{HTML}{FD8800}
\definecolor{dark_red}{rgb}{0.7, 0., 0.}
\definecolor{light_pink}{rgb}{1,0.4,0.4}
\definecolor{lblue}{rgb}{0.384602,0.117763,0.973947}
\usepackage{color}
\input{colordvi.tex}
\allowdisplaybreaks[1]
\usepackage{url}
\usepackage{hyperref}
\hypersetup{
colorlinks=false,
hidelinks}

\newcommand{\bk}{\mathbf{k}}

\newcommand*{\D}{{\rm d}}

\definecolor{MONZA}{HTML}{CF000F}
\definecolor{DARKBLUE}{HTML}{00008b}
\definecolor{DARKMAGENTA}{HTML}{8b008b}
\definecolor{DARKCYAN}{HTML}{00cfc0}

\begin{document}

\title{Scalar parity-odd trispectrum from gravitational Chern-Simons interaction vertices}

\author[a]{Giorgio Orlando,}
\author[a]{Shingo Akama,}
\author[a]{Chunshan Lin}


\affiliation[a]{Faculty of Physics, Astronomy and Applied Computer Science, Jagiellonian University, 30-348 Krakow, Poland}

\emailAdd{giorgio.orlando@uj.edu.pl, shingo.akama@uj.edu.pl, chunshan.lin@uj.edu.pl}

\date{\today}

\abstract{
In this paper, we explore parity violation in a scalar trispectrum from a dynamical Chern-Simons gravity theory.
So far, a graviton-mediated diagram with two vertexes being of general relativity has been studied in this theory by taking into account the impact of a modified dispersion relation of gravitons on graviton’s bulk propagators. 
We instead study a parity-odd trispectrum from both a graviton-mediated diagram, where one of the two vertexes originates from the Chern-Simons term, and a contact diagram by using the bulk propagators in general relativity. 
After computing the scalar-scalar-tensor cubic interactions and the scalar quartic ones originating from the Chern-Simons term, first we show that the resultant parity-odd trispectrum vanishes in the case of Bunch-Davies initial conditions, which is consistent with a no-go theorem for a non-vanishing parity-odd trispectrum. Then, we discuss a way to acquire a non-vanishing parity-odd trispectrum from the viewpoint of non-Bunch-Davies initial conditions.}

\maketitle
\flushbottom

\section{Introduction}

The study of primordial non-Gaussianities has emerged as a powerful tool for probing the physics of the early universe, particularly in the context of inflationary scenarios beyond the minimal slow-roll paradigm (see e.g. \cite{Achucarro:2022qrl}). In this context, parity-violating signatures in primordial correlation functions have attracted significant attention, as they provide a unique observational window into high-energy physics, including potential extensions of general relativity (GR). From an observational standpoint, an evidence of parity-violation in the BOSS-galaxies trispectrum was first reported in~\cite{Hou:2022wfj,Philcox:2022hkh}, potentially pointing to phenomena rooted in the early universe. However, a more recent analysis of the CMB trispectrum from \textit{Planck} data finds no evidence for parity-violation from inflation~\cite{Philcox:2023ypl}. Nonetheless, ongoing improvements in the volume and quality of CMB data might eventually lead to a net detection in the future. Among parity-violating theories, those incorporating gravitational Chern-Simons (gCS) terms are particularly noteworthy due to their intrinsic violation of parity and their emergence in various ultraviolet (UV)-motivated frameworks such as string theory and effective field theory approaches to quantum gravity (see e.g. \cite{Lue:1998mq,Alexander:2004wk,Satoh:2010ep,Dyda:2012rj,Bartolo:2017szm,Creque-Sarbinowski:2023wmb,Christodoulidis:2024ric,Orlando:2025pkb} for  related works in the context of inflation).

While previous studies have primarily focused on the bispectrum level (see e.g. \cite{Bartolo:2017szm}), where parity-violating signals from gCS terms can be enhanced under specific conditions, such as non-Bunch-Davies (non-BD) initial states (see e.g. \cite{Gong:2023kpe,Christodoulidis:2024ric}), relatively less is known about their imprint on the scalar trispectrum. Notably, it has been established that under Bunch-Davies initial conditions and massless fields in the interaction vertices, parity-odd contributions to the scalar trispectrum vanish due to a no-go theorem, barring logarithmic infrared divergences in the in-in time-integrations (see e.g. \cite{Liu:2019fag,Cabass:2022rhr,Stefanyszyn:2023qov,Thavanesan:2025kyc}). Earlier investigations within dynamical Chern-Simons gravity considered only graviton-mediated diagrams with both vertices sourced from GR, where parity violation was encoded in the modified tensor mode function~\cite{Creque-Sarbinowski:2023wmb}.

In this work, we extend the current understanding by computing the scalar parity-odd trispectrum sourced by Chern-Simons interaction vertices. Specifically, we analyze both graviton-mediated and contact-interaction diagrams, where at least one vertex arises from the Chern-Simons term. We employ the standard GR bulk graviton propagators and derive the scalar-scalar-tensor cubic and scalar quartic interactions relevant to these processes. We first confirm that, in agreement with the no-go theorem, the parity-odd trispectrum vanishes under Bunch-Davies initial conditions. This is due to the absence of infra-red divergences carried by the new interaction vertices computed in this paper (Eqs. \eqref{eq:sst_CS_new} and \eqref{eq:2orderCS}, \eqref{eq:newGR_odd}). We then go beyond this result by introducing non-Bunch-Davies initial states, demonstrating that such configurations can give rise to a non-vanishing parity-odd scalar trispectrum.

The paper is organized as follows: Sec. \ref{sec:2} lays out the theoretical framework by defining the Pontryagin density and introducing the gCS term in the context of slow-roll inflation. This section also specifies the gauge choice, perturbation variables, and background dynamics relevant for our analysis. In Sec. \ref{sec:3}, we review the quadratic theory for tensor modes, showing how the gCS term modifies their dynamics. In Sec. \ref{sec:4} we compute the scalar-scalar-tensor cubic and scalar quartic interaction vertices. Sec. \ref{sec:5} is dedicated to the main computational results: in Sec. 5.1, we calculate the scalar parity-odd trispectrum from graviton-mediated diagrams involving a Chern-Simons vertex. In Sec. 5.2, we compute the contribution from contact interaction diagrams. Sec. \ref{sec:6} investigates scenarios beyond Bunch-Davies initial conditions. Finally, Sec. \ref{sec:7} provides a combined summary and conclusions.

\section{Basics and Pontryagin Density Definition}
\label{sec:2}

The standard action during (slow-roll) inflation reads
\begin{align}
S_{\rm slow-roll} = \int d^4 x \sqrt{-g} \, \left[ \frac{M^2_{pl}}{2} R - \frac{1}{2}(\partial_\mu \phi)(\partial^\mu \phi) - V(\phi) \right] \, , 
\end{align}
where $M_{pl}$ is the reduced Planck Mass. In our paper, we study parity-violating signatures of the perturbations around a spatially flat Friedmann-Lema\^{i}tre-Robertson-Walker (FLRW) background metric
\begin{align}
\D s^2&=-\D t^2+a^2(t)\delta_{ij}\D x^i \D x^j\notag\\
&=a^2(\tau)(-\D\tau^2+\delta_{ij}\D x^i \D x^j) \, ,
\end{align}
with $a$ the Roberson-Walker scale factor. In the second line, we introduced the conformal time $\tau$.
The background equations of motion in the FLRW spacetime read
\begin{align}
3 M_{pl}^2H^2 &= \frac{1}{2} \dot \phi^2 + V(\phi) \, , \\
M_{pl}^2\dot H &= -\frac{1}{2} \dot \phi^2 \, , \\ 
0 &= \ddot \phi + 3 H \dot \phi + \partial_\phi V(\phi) \, , 
\end{align}
where $H = \dot a/a$. The dot means differentiation with respect to $t$, while the prime will mean differentiation with respect to  $\tau$. The slow-roll parameters are defined as
\begin{align}
\epsilon &= \frac{1}{2} \left(\frac{M_{pl} \, \partial_\phi V}{V}\right)^2 \simeq \frac{1}{2} \frac{\dot \phi^2}{H^2} \frac{1}{M^2_{pl}} \, , \\
\eta &=  \frac{M^2_{pl} \, \partial_{\phi\phi}V}{V} \simeq - \frac{\ddot \phi}{\dot \phi H} + \frac{1}{2} \frac{\dot \phi^2}{H^2} \frac{1}{M^2_{pl}} \, . 
\end{align}
The gravitational Chern-Simons operator (also known as Pontryagin density) can be expressed as the divergence of the gravitational Chern-Simons current $K^{\mu}$ as~\cite{Jackiw:2003pm}
\begin{align}
\nabla_\mu K^\mu &=  \frac{1}{2}\, ^*{R} R  \, ,\\
K^\mu &= \epsilon^{\mu\nu\alpha\beta}\left(\Gamma^\sigma_{\nu\rho}\partial_\alpha \Gamma^\rho_{\beta\sigma} + \frac{2}{3}\Gamma^\sigma_{\nu\rho}\Gamma^\rho_{\alpha\lambda}\Gamma^\lambda_{\beta\sigma}\right) \, ,
\label{eq:4current}
\end{align}
where $^*{R} R$ is defined by the dual Riemann tensor as
\begin{align}
^*{R} R :=\frac{1}{2}\epsilon^{\mu\nu\alpha\beta}{R^\rho}_{\sigma\alpha\beta}{R^\sigma}_{\rho\mu\nu},
\end{align}
which can be written as
\begin{equation} \label{eq:pot}
^*{R} R = 2 \, \nabla_\mu K^\mu = 2 \partial_\mu K^\mu + 2 \Gamma_{\mu \delta}^\mu K^\delta \, .
\end{equation}
We consider the Pontryagin term during inflation coupled to the inflaton field $\phi$ through a generic coupling $f(\phi)$. This coupling is necessary, otherwise the theory is trivial since the Chern-Simons term is a topological one. The resulting dynamical Chern-Simons contribution reads
\begin{align} \label{eq:CS}
\Delta S_{CS} = \int d^4 x \sqrt{-g} \, f(\phi) ^*{R} R &= 2 \int d^4 x \sqrt{-g} \, f(\phi) \nabla_\mu K^\mu \nonumber \\
&=2 \int d^4 x  \,f(\phi) \partial_\mu (\sqrt{-g} K^\mu) = - 2 \int d^4 x \sqrt{-g}  \, (f'(\phi)   K^0 + \partial_i f(\phi) \, K^i) \, .
\end{align}
The total action of our theory is
\begin{equation} 
S = S_{\rm slow-roll} + \Delta S_{CS} \, .
\end{equation}
The additional Chern-Simons contribution does not act on the background equations, but it does affect primordial perturbations. At quadratic level, it does affect tensor perturbations only, while the scalar sector is affected only starting from the trispectrum. 

To perform our computations of the scalar trispectrum, we fix the spatially flat gauge, where the perturbed metric is of the form  (see e.g. \cite{Maldacena:2002vr})
\begin{align}
ds^2 &= g_{\mu\nu} dx^\mu dx^\nu = -(N^2 - N^i N_i) \, a^2 d\tau^2 + 2 N_i \, a \, dx^i d\tau + g_{ij} \, dx^i dx^j \, ,
\end{align}
where
\begin{align} 
N=1+\alpha \, , \quad \quad  N_i&=\partial_i\theta + \beta_i \, ,\quad \quad  g_{ij} = a^2 \exp[h]_{ij} \, , \nonumber \\
 \partial_i \beta^i &= 0 \, , \qquad\qquad \partial^i h_{ij}= 0 \, , h_{i}^i= 0 \, ,
\end{align}
with $\alpha$, $\theta$ and $\beta_i$ auxiliary fields, and 
\begin{align}
\exp[h]_{ij}=\delta_{ij}+h_{ij}+\frac{1}{2}h_{ik}h^k_j+\cdots \, , 
\end{align}
with $h_{ij}$ tensor perturbations. The inflaton is perturbed as 
\begin{align}
\phi=\phi_0(t)+\delta\phi \, .
\end{align}
This gauge is convenient in order to identify the terms leading in slow-roll. Also, as we will see in Sec.~\ref{sec:4}, this allows us to compute the relevant terms limiting the amount of perturbative expansion of the metric components. 

For the purpose of the computations below, it is helpful to remind the zero-th order equation of motions for the scalar and tensor perturbations in the standard GR
\begin{align}
\delta\phi'' + 2 a H \delta\phi' - \partial^2 \delta\phi = 0 \, , \\ 
h_{ij}'' + 2 a H h_{ij}' - \partial^2 h_{ij} = 0 \, .
\end{align}
Later on we will be interested in computing the parity-odd scalar trispectrum due to the additional interaction terms introduced by the Chern-Simons operator. Once expanded in perturbations and computed the cosmological correlators in the spatially flat gauge in terms of the inflaton perturbations $\delta\phi$, the connection with the so-called gauge-invariant curvature perturbations $\zeta$ can be done by virtue of the Wick theorem after specifying the nonlinear transformation that connects the spatially flat and co-moving gauges on super-horizon scales. This transformation is of the form \cite{Maldacena:2002vr}
\begin{equation}
\zeta = b_1 \delta\phi + b_2 \delta\phi^2\, , 
\end{equation}
where
\begin{equation}
b_1 = - \frac{1}{\sqrt{2 \epsilon} M_{pl}} \, , 
\end{equation}
and $b_2 \delta\phi^2$ is a schematic way to write all the terms of quadratic order in $\delta\phi$, e.g., $\delta\phi^2, \delta\phi\dot{\delta\phi}$, etc., see Ref.~\cite{Maldacena:2002vr} for the explicit expressions.
Therefore, the scalar trispectrum of $\zeta$ can be related to that of $\delta\phi$ by the schematic relation
\begin{equation}
\langle\zeta^4\rangle \sim b_1^4 \langle\delta\phi^4\rangle + b_1^3 b_2 \, \langle\delta\phi^3\rangle \langle\delta\phi^2\rangle + b_1^2 b_2^2 \,\langle\delta\phi^2\rangle \langle\delta\phi^2\rangle \langle\delta\phi^2\rangle + \cdots \, ,
\end{equation}
where the ellipsis indicates terms of higher order in the perturbative expansion and slow-roll.
From this equation, it follows that when we take the parity-odd part of the scalar trispectrum, we get the linear relation 
\begin{equation} \label{eq:phi_to_zeta4}
\langle\zeta^4\rangle_{P.O.} = b_1^4 \, \langle\delta\phi^4\rangle_{P.O.}  \, ,
\end{equation}
since the scalar power spectrum and the scalar auto-bispectrum are insensitive to parity violation (see e.g. \cite{Shiraishi:2016mok}). This is not surprising as the non-linear transformation that allows us to switch from the spatially flat gauge to the comoving one is insensitive to parity transformation. For what said, in the following we can compute the parity-odd scalar trispectrum of $\delta\phi$ and obtain the parity-odd scalar trispectrum of $\zeta$ using the linear relation \eqref{eq:phi_to_zeta4}. 

\section{Review of a quadratic theory} \label{sec:3}
The quadratic action for gravitons originating from the standard GR action reads
\begin{equation}
S^{(2)}_{GR} = \frac{M^2_{pl}}{8} \int d^4 x \, a^2 \left[ (h_{ij})' (h^{ij})' + (\partial^2 h_{ij}) (h^{ij}) \right] \, ,
\end{equation}
where Latin contractions are made with the Kronecker Delta, and we adopted the conformal time. At quadratic level, only the tensor modes receive a correction from the Chern-Simons term. By implementing the correction given in Eq. \eqref{eq:CS}, the full quadratic action for gravitons in the dynamical Chern-Simons theory reads \cite{Creque-Sarbinowski:2023wmb}
\begin{align} 
S^{(2)} &= S^{(2)}_{GR} + \Delta S^{(2)}_{CS} \nonumber \\
&= \frac{M^2_{pl}}{8} \int d^4 x \, a^2 \Big\{\left[(h_{ij})' (h^{ij})' + (\partial^2 h_{ij}) (h^{ij}) \right] \nonumber \\
&\quad\ - 8 \, a^{-1} \, \partial_\phi f(\phi) \, \sqrt{2 \epsilon} \, \frac{H}{M_{pl}} \,  \bar \epsilon^{ijk} \left[h'_{il} \partial_j h'_{lk} +  (\partial^2 h_{il}) \partial_j h_{lk} \right] \Big\} \, .
\end{align} 
It is convenient to define the Chern-Simons mass as
\begin{equation}
M_{CS} = \left[8 \, \left|\partial_\phi f(\phi)\right| \, \sqrt{2 \epsilon} \, \frac{H}{M_{pl}}\right]^{-1} \, .
\end{equation}
Then, the previous equation can be rewritten as\footnote{Here we assume $\partial_\phi f(\phi)>0$. In case $\partial_\phi f(\phi)<0$, a sign switch is introduced which results in the exchange of R-handed tensor perturbations with L-handed ones in the final equations.} 
\begin{align} \label{eq:action_modified}
S^{(2)} = \frac{M^2_{pl}}{8} \int d^4 x \, a^2 \Big\{\left[(h_{ij})' (h^{ij})' + (\partial^2 h_{ij}) (h^{ij}) \right] - \, a^{-1} \, M^{-1}_{CS} \,\,\, \bar \epsilon^{ijk} \left[h'_{il} \partial_j h'_{lk} +  (\partial^2 h_{il}) \partial_j h_{lk} \right] \Big\} \, .
\end{align}
We now go to the Fourier space. For this purpose, we write the tensor perturbation as 
\begin{equation}
h_{ij}(\tau, \mathbf x) = \sum_{\lambda = R/L} \int\frac{d^3 \mathbf k}{(2 \pi)^3} \, \left[ h_\lambda (\tau,{\mathbf k}) \, e^{\lambda}_{ij}(\mathbf k)  \right] \, e^{i \mathbf{k} \cdot \mathbf{x}} \, , 
\end{equation}
where we have employed the decomposition in terms of helicity (or chiral) states $\lambda = R, L$. For our purposes, we express tensor perturbations using the chiral (helicity) polarization basis, defined as
\begin{align}
e_{ij}^{R, L} &= \frac{1}{\sqrt{2}} \left( e_{ij}^+ \pm i e_{ij}^\times \right) \, , \\
h_{R, L} &= \frac{1}{\sqrt{2}} \left( h_+ \mp i h_\times \right) \, ,
\end{align}
where $h_{+,\times}$ and $e_{ij}^{+, \times}$ denote the standard linear polarizations of the tensor modes.

We recall that, given a tensor wavevector expressed in spherical coordinates as
\begin{equation}
\hat{k} := \frac{\mathbf{k}}{|\mathbf{k}|} = (\sin\theta\cos\phi, \sin\theta\sin\phi, \cos\theta) \, ,
\end{equation}
the linear polarization tensors can be constructed using two orthonormal vectors orthogonal to $\hat{k}$,
\begin{align}
e_{ij}^{+} &= (u_1)_i (u_1)_j - (u_2)_i (u_2)_j \,, \\
e_{ij}^{\times} &= (u_1)_i (u_2)_j + (u_2)_i (u_1)_j \,,
\end{align}
with
\begin{align}
u_1 &= (\sin \phi , -\cos \phi, 0) \,, \\
u_2 &=
\begin{cases}
(\cos \theta \cos \phi, \cos \theta \sin \phi, -\sin \theta) & \text{for } \theta \leq \pi/2 \, , \\ 
-(\cos \theta \cos \phi, \cos \theta \sin \phi, -\sin \theta) & \text{for } \theta > \pi/2 \, .
\end{cases}
\end{align}
The chiral polarization tensors satisfy the following normalization and symmetry identities:
\begin{align}
e_{ij}^{L}(\hat{k}) e_{L}^{ij}(\hat{k}) &= e_{ij}^{R}(\hat{k}) e_{R}^{ij}(\hat{k}) = 0 \,, \nonumber \\
e_{ij}^{L}(\hat{k}) e_{R}^{ij}(\hat{k}) &= 2 \, , \nonumber \\
e_{ij}^{\lambda}(-\hat{k}) &= e_{ij}^{\lambda}(\hat{k}) \,, \nonumber \\
e_{ij}^{R*}(\hat{k}) &= e_{ij}^{L}(\hat{k}) \,, \nonumber \\
h^{R*}_{-\mathbf{k}} &= h^L_{\mathbf{k}} \, , \nonumber \\
k_l \, \bar{\epsilon}^{mlj} \, e_j^{(\lambda) i}(\hat{\mathbf{k}}) &= \mp i \alpha_\lambda \, k \, e^{(\lambda) im}(\hat{\mathbf{k}}) \,, \label{eq:circ_identities}
\end{align}
where $\alpha_R=+1$ and $\alpha_L=-1$, and $\bar \epsilon^{mlj}$ denotes the Levi-Civita anti-symmetric symbol. The upper ($-$) or lower ($+$) sign in the last identity corresponds to $\theta \leq \pi/2$ and $\theta> \pi/2$, respectively. In the following, we assume $\theta \leq \pi/2$, though the final results are insensitive to this choice.

With these conventions, the modified action in Eq.~\eqref{eq:action_modified} becomes
\begin{align}  \label{eq:action_modified2}
S^{(2)} = \frac{M^2_{pl}}{4} \sum_{\lambda = R/L}  \int d \tau \, \int \frac{d^3 \bk}{(2 \pi)^3}  \, A_\lambda^2 \Big\{ \, |h_{\lambda}'|^2 - k^2 |h_{\lambda}|^2  \, \Big\} \, ,
\end{align}
where
\begin{equation}
A_\lambda^2 = a^2 \left[ 1 - \alpha_\lambda \frac{k}{a \, M_{CS}} \right]  = a^2 \left[ 1 - \alpha_\lambda \frac{k}{a H}\frac{H}{M_{CS}} \right] \, .
\end{equation}
Defining $a_o$ as the scale factor at the beginning of inflation, it is evident that the R-handed modes become ghost-like when their co-moving momenta are larger than
\begin{equation}
k_{CS} = a_o H \, \frac{M_{CS}}{H} \, .
\end{equation}
Therefore, we need to introduce an UV-cutoff scale to the theory below this $k_{CS}$ energy scale. We note that if $H/M_{CS}>1$, then $k_{CS} < a_o H< a_H H$, with $a_H$ denoting the scale factor at the horizon crossing. It follows that the theory breaks down for all the modes inside the horizon. Thus, we must impose $H/M_{CS} \ll 1$ in order to have physical modes for which the theory is not spoiled within the horizon\footnote{We note that in some other references that study the gravitational Chern-Simons term during inflation, like \cite{Creque-Sarbinowski:2023wmb}, a more relaxed condition $H/M_{CS} < 1$ is imposed. While $H/M_{CS} < 1$ is enough to get some modes for which the theory makes sense, as $H$ approaches $M_{CS}$ other terms proportional to higher powers of $H/M_{CS}$ start to be relevant in the equation of motion \eqref{eq:eom_CS}, resulting in increasing departures of the real solution with the solution given in Eq. \eqref{eq:CS_solution_graviton}.}.   

By performing the following field re-definition
\begin{equation}
\tilde h_\lambda = A_\lambda \, h_\lambda \, ,
\end{equation}
the previous action reads
\begin{align}  \label{eq:action_modified3}
S^{(2)} = \frac{M^2_{pl}}{4} \sum_{\lambda = R/L}  \int d \tau \, \int \frac{d^3 \bk}{(2 \pi)^3}  \, \Big\{ \, |\tilde h_{\lambda}'|^2 - k^2 |
\tilde h_{\lambda}|^2  + \frac{A''_\lambda}{A_\lambda}|
\tilde h_{\lambda}|^2 \, \Big\} \, .
\end{align}
Up to the leading-order in both the ratio $H/M_{CS}$ and slow-roll, the effective mass term reads\footnote{Here we have ignored terms proportional to $\dot H$ as they are not relevant for the final mode-function.}
\begin{equation}
\frac{A''_\lambda}{A_\lambda} = \frac{2}{\tau^2} - \alpha_\lambda \frac{k}{\tau}\frac{H}{M_{CS}}  \, .
\end{equation}
Therefore, the equation of motion for the canonically normalized field $\tilde h_\lambda$ reads
\begin{equation} \label{eq:eom_CS}
\tilde h''_\lambda + \left(k^2 - \frac{2}{\tau^2} + \alpha_\lambda \frac{k}{\tau}\frac{H}{M_{CS}} \right) \tilde h_\lambda = 0 \, .
\end{equation}
As usual, we canonically quantize this field as
\begin{equation}
\tilde h_\lambda(\tau, \mathbf x) = \, \left[ u_\lambda (\tau,{\mathbf k}) \, \hat b_\lambda(\mathbf k) + u^*_\lambda (\tau,-{\mathbf k}) \, \hat b^{\dagger}_\lambda(-\mathbf k) \right] \, . 
\end{equation}
By imposing the Bunch-Davies (BD) initial condition, we get the following solution for the mode function of the original field $h_\lambda$ (see e.g. \cite{Creque-Sarbinowski:2023wmb})
\begin{align} \label{eq:CS_solution_graviton}
A^{-1}_{R/L} \, u_{R/L} (\tau, k) &=  \left[1 \mp \frac{k}{a H}\frac{H}{M_{CS}} \right]^{-1/2} \frac{H}{M_{pl} \sqrt{k^3}} \, (1+ i k \tau) \, e^{- i k \tau} \, \left[\frac{U(2 \pm i \frac{H}{2 M_{CS}}, 4, 2 i k \tau)}{U(2, 4, 2 i k \tau)}\right]  \notag\\
&\quad\ \times \exp\left(\pm\frac{\pi}{4}\frac{H}{M_{CS}}\right),
\end{align}
where U is the confluent hypergeometric function of the second kind.

On super-horizon scales defined by $-k\tau\ll1$, the physical effect of the Chern-Simons term solely comes from the exponential. For $H/M_{CS} \ll 1$, the amount of circular polatrization is predicted to be
\begin{equation}
\Pi = \frac{P_R - P_L}{P_R + P_L} = \frac{\pi}{2} \frac{H}{M_{CS}} \ll 1 \, ,
\end{equation}
where $P_{R/L}$ denotes the super-horizon tensor power spectrum of each polarization mode. Natural values for $\Pi$ are thus very small. 

Below we will consider the parity-violation generated in the scalar trispectrum by the interaction terms between two scalars and a graviton as well as the self interaction of four scalars. Conversely, we expand the tensor mode function in slow-roll and in power series of $H/M_{CS}$ and take only the 0-th order term, which is parity invariant. This is equivalent to imposing 
\begin{equation} \label{eq:mode_function}
u_\lambda (\tau, k) =  \frac{H}{M_{pl} \sqrt{k^3}} \, (1+ i k \tau) \, e^{- i k \tau} \, .
\end{equation}
Regarding the scalar sector, at quadratic level this is left untouched with respect to standard slow-roll inflation. The corresponding scalar mode-function with BD initial condition reads  
\begin{align} \label{eq:scalar_BD}
u_{\delta \phi}(\tau, k)=  \frac{H}{\sqrt{2 k^3}}(1+ik\tau) \, e^{-ik\tau} \, .
\end{align}

\section{Cubic and quartic actions} \label{sec:4}
In this section, let us show the explicit expressions of the scalar-scalar-tensor and scalar quartic interactions which are necessary to compute the contributions to the parity-odd scalar trispectrum from the graviton-exchange and contact diagrams, respectively. For simplicity of notation, in the following we will take $M_{pl} = 1$.

\subsection{Scalar-scalar-tensor cubic action}
\label{Appendix: perturbed actions_sst}

In the spatially-flat gauge, primordial non-Gaussianities from Eq. \eqref{eq:CS} originate by the Taylor expanded coupling as 
\begin{align} \label{eq:CS2}
\Delta S_{CS} = - 2 (\partial_\phi f) \int d^4 x \sqrt{-g}  \, [\delta\phi'   K^0 + (\partial_i \delta\phi) \, K^i] \, .
\end{align}
Let us first consider the perturbed FLRW metric of the form
\begin{align}
ds^2 &= g_{\mu\nu} dx^\mu dx^\nu = -(N^2 - N^i N_i) \, a^2 d\tau^2 + 2 N_i \, a \, dx^i d\tau + a^2 \, \tilde g_{ij} \, dx^i dx^j \, ,
\end{align}
where we have factorized the scale-factor out of the metric as $\tilde g_{ij} = g_{ij}/a^2$. 
By redefining $N_i = a \tilde N_i$, we get 
\begin{align}
ds^2 &= g_{\mu\nu} dx^\mu dx^\nu = -(N^2 - a^2 \tilde N^i \tilde N_i) \, a^2 d\tau^2 + 2 \tilde N_i \, a^2 dx^i d\tau + a^2 \, \tilde g_{ij} \, dx^i dx^j \, .
\end{align}
Exploiting the fact that the Pontryagin density is invariant under a conformal transformation of the metric \cite{Tian:2015vda}, we can compute it after making the conformal transformation $g'_{\mu \nu} = a^{-2} \, g_{\mu \nu}$, leading to the transformed metric 
\begin{align} \label{eq:direct_metric_CS}
ds^2 &= g'_{\mu\nu} dx^\mu dx^\nu = -(N^2 - a^2 \tilde N^i \tilde N_i) \,  d\tau^2 + 2 \tilde N_i \, dx^i d\tau + \, \tilde g_{ij} \, dx^i dx^j \, .
\end{align}
The above can be written in terms of the inverse metric as
\begin{align}
ds^2 & = g'^{\mu\nu} dx_\mu dx_\nu = - N^{-2} \, d\tau^2 + 2 a^2  \tilde N^i N^{-2}  \, dx_i d \tau + \, \left(\tilde g^{ij}- a^2\frac{\tilde N^i \tilde N^j}{N^2}\right) \, dx_i dx_j \, .
\end{align}
From now on, we will omit the prime which indicates the transformed metric. The Christoffel symbols $\Gamma^\mu_{\alpha\beta}$ read
\begin{equation}
\Gamma^\mu_{\alpha\beta} = \frac{g^{\mu\lambda}}{2}\left(\partial_\alpha g_{\lambda\beta} + \partial_\beta g_{\alpha\lambda }  - \partial_\lambda g_{\alpha\beta}\right) \, .
\end{equation}
As we will see better later on, for our computations we need to compute the Christoffel symbols only up to first order in the perturbations:
\begin{align}
{\Gamma^\mu_{\alpha\beta}} = &\frac{\left[g^{\mu\lambda}\right]^{(0)}}{2} \left[\left(\partial_\alpha g_{\lambda\beta} + \partial_\beta g_{\alpha\lambda}  - \partial_\lambda g_{\alpha\beta}\right)\right]^{(1)} + \mbox{(quadratic)} \, ,
\end{align}
where the suffixes refer to the order of each term in perturbation theory. Note that since in the transformed background metric the scale factor disappeared, the Christoffel symbols are vanishing on the background, $\Gamma^{(0)} = 0$. By expanding in perturbations the transformed metric components up to the desired order, we get:
\begin{align} \label{eq:metric}
 & g_{00} = -1 - 2 N  ...\, , \qquad g_{0i} = g_{i0} = \tilde N_i \, , \qquad  g_{ij} = \delta_{ij} + h_{ij} + ... \, , \nonumber \\
 & g^{00} = -1  ...\, , \qquad g^{0i} = g^{i0} =  0 + ... \, , \qquad  g^{ij} = \delta_{ij}  + ... \, ,
\end{align}
where $N$ here stands for the perturbed part of the lapse function only. This allows to compute the Christoffel symbols making Latin-contractions with the Kronecker delta. 

By expanding the Christoffel symbols up to quadratic order, we obtain 
\begin{align}
\Gamma^0_{00} &= N' + \mbox{(quadratic)} \, ,\\
\Gamma^0_{0i} &= \partial_i N  + \mbox{(quadratic)} \, ,\\
\Gamma^{0}_{ij} &= -\frac{1}{2} \partial_i \tilde N_j - \frac{1}{2}\partial_j \tilde N_i + \frac{1}{2} h_{ij}' + \mbox{(quadratic)} \, ,  \\
\Gamma^i_{00} &= \partial_i N + {\tilde N_i}'  + \mbox{(quadratic)} \, ,  \\
\Gamma^i_{0j} &=  \frac{1}{2} \partial_j \tilde N_i - \frac{1}{2} \partial_i \tilde N_j + \frac{1}{2} h'_{ij}  + \mbox{(quadratic)} \, , \\
\Gamma^i_{jk} &= \frac{1}{2} \partial_k h_{ij} + \frac{1}{2} \partial_j h_{ik} - \frac{1}{2} \partial_i h_{jk} +  \mbox{(quadratic)} \, .
\end{align}
Using these Christoffel symbols, in the following we compute the Chern-Simons 4-current \eqref{eq:4current}. Given the structure of Eq. \eqref{eq:CS2}, we note that we only need the 4-current up to quadratic order. Since $\Gamma^{(0)} = 0$, only the first order Christoffel symbols are needed. Then, the schematic form of the 4-current is
\begin{align} 
K^{\mu \, (2)}  \sim \Gamma^{(1)}  \partial\Gamma^{(1)} \, . 
\end{align} 
In the next, we perform the computation in detail
\begin{align} \label{eq:kappa0_current}
K^0 
&= \epsilon^{ijk}\Bigg{\{}  \frac{1}{2}\left (\partial_l \tilde N_i\right) \left(\partial_j\partial_l \tilde N_k\right) - \frac{1}{2}\left (\partial_l \tilde N_i\right) \left(\partial_j h'_{lk}\right) - \frac{1}{2} h'_{il} \left(\partial_j\partial_l \tilde N_k\right) + \frac{1}{2} h'_{il}\left(\partial_j h'_{lk}\right)   \nonumber \\
& \qquad\qquad + \frac{1}{2} \left(\partial_m h_{il}\right) \left(\partial_j\partial_l h_{km}\right)- \frac{1}{2} \left(\partial_m h_{il}\right) \left(\partial_j\partial_m h_{kl}\right) \Bigg{\}} + \mbox{(cubic)} \, .\\
K^i &= \epsilon^{ijk}\Bigg{\{}  \frac{1}{2} \left(\partial_l \tilde N_j\right) \left(\partial_l \tilde N'_k\right) - \frac{1}{2} \left(\partial_l \tilde N_j\right) h''_{lk} - \frac{1}{2}  h'_{lj} \left(\partial_l \tilde N'_k\right) +  \frac{1}{2}  h'_{jl} h''_{lk} + \frac{1}{2} \left(\partial_m h_{jl}\right) \left(\partial_l h'_{km}\right)   \nonumber \\
&\qquad\qquad - \frac{1}{2} \left(\partial_m h_{jl}\right) \left(\partial_m h'_{lk}\right) + 2 \left(\partial_l N\right) \left(\partial_l \partial_j \tilde N_k\right) - 2 \left(\partial_l N\right) \left(\partial_j h'_{lk}\right) +  \tilde N'_l \left(\partial_l \partial_j \tilde N_k \right) -  \tilde N'_l \left(\partial_j h'_{lk}\right)  \nonumber \\
& \qquad\qquad - \left(\partial_m \tilde N_l\right) \left(\partial_j \partial_l h_{km}\right) +  \left(\partial_m \tilde N_l\right) \left(\partial_j \partial_m h_{kl}\right)\Bigg{\}} + \mbox{(cubic)} \, . \label{eq:kappai_current}
\end{align}
By reminding the relation between the covariant Levi-Civita tensor and the Levi-Civita symbol $\bar \epsilon^{ijk}$:
\begin{align}
\bar \epsilon^{ijk} = \sqrt{-g} \, \epsilon^{ijk} \, , 
\end{align}
the final cubic Lagrangian can be obtained by substituting the 4-current in Eq. \eqref{eq:CS2} with $\bar \epsilon^{ijk}$ replacing the covariant Levi-Civita tensor in Eqs. \eqref{eq:kappa0_current} and \eqref{eq:kappai_current}. 

Doing this computation and isolating the \textit{sst} terms, we get the following Lagrangian at leading order in slow-roll
\begin{align} \label{eq:Lsst_flat}
\left[\Delta \mathcal L_{CS}\right]^{(3)sst} = \, & - 2 \, (\partial_\phi f) \, \bar\epsilon^{ijk} \left[  - \frac{1}{2} \delta \phi'  \left (\partial_l \tilde N_i\right) \left(\partial_j h'_{lk}\right) - \frac{1}{2}  (\partial_i \delta \phi) \left(\partial_l \tilde N_j\right) h''_{lk} - \frac{1}{2} (\partial_i \delta \phi)  h'_{lj} \left(\partial_l \tilde N'_k\right) \right. \nonumber \\
&\left. \qquad \qquad \qquad \qquad  -  (\partial_i \delta \phi) \, \tilde N'_l \left(\partial_j h'_{lk}\right)  - 2 \, (\partial_i \delta \phi)  \left(\partial_l N\right) \left(\partial_j h'_{lk}\right) \right] \, . 
\end{align}
At this point, it is useful to remind the first order values of the auxiliary fields $N$ and $N_i$ in terms of the inflaton perturbation in the spatially flat gauge \cite{Maldacena:2002vr} \footnote{Note the additional factor $a$ in the $N_i$ computation of Maldacena as here we are using the rescaled constraint $\tilde N_i = N_i/a$.} (these are valid assuming slow-roll phase):
\begin{align} \label{eq:constraints}
N = \sqrt{\frac{\epsilon}{2}} \, \delta \phi \, , \qquad 
\tilde N_i = \epsilon \, \partial_i \, \partial^{-2} \, \frac{d}{d \tau} \Big[ - \frac{H}{\dot \phi} \, \delta \phi \Big] \simeq - \sqrt{\frac{\epsilon}{2}} \, (\partial_i \, \partial^{-2} \,\delta \phi') \, .
\end{align}
After using the zero-th order equations of motion to eliminate second order derivatives in Eq. \eqref{eq:Lsst_flat} and substituting in the Lagrangian the constraint equations \eqref{eq:constraints} (and doing appropriate integrations by parts), we get
\begin{align} \label{eq:sst_CS_new}
\left[\Delta \mathcal L_{CS}\right]^{(3)sst, \, flat} =&   4 \sqrt 2 \,  \epsilon^{1/2} \, (\partial_\phi f) \, a H \, \bar\epsilon^{ijk} \, \left[ (\partial_i \delta \phi) \left(\partial_l  \partial^{-2} \delta \phi' \right) \left(\partial_j h'_{lk}\right)   \right] \nonumber\\
=& \frac{1}{2 M_{CS}} a \, \bar\epsilon^{ijk} \, \left[ (\partial_i \delta \phi) \left(\partial_l  \partial^{-2} \delta \phi' \right) \left(\partial_j h'_{lk}\right)   \right] \, .
\end{align}
We find some discrepancy with respect to Eq. 5.17 of~\cite{Bartolo:2017szm} which first computed this interaction using the direct contraction between two Weyl tensors and the Levi-Civita tensor. While the overall time-scaling matches, here we get an additional inverse Laplacian applied to one of the scalars. We argue that the result in the current publication is more reliable as it was obtained via linear quantities using the conformal transformation trick, while the result in Ref.~\cite{Bartolo:2017szm} follows from the computation of the Weyl tensor at non-linear order.

\subsection{Scalar quartic action} \label{sec:4_quartic}
We then move on to the computations of the scalar quartic terms. To do so, it is convenient to set $h_{ij}=0$ in the metric tensor \eqref{eq:direct_metric_CS}. As we will see below, the parity-odd quartic action comes from either $S_{\rm CS}$ and $S_{\rm slow-roll}$. In the following we start computing the contribution of the former.

By looking to Eq. \eqref{eq:CS}, we infer that the quartic Lagrangian contributed by the gCS gravitational term comes from
\begin{align}  
\Delta S_{CS} = - 2( \partial_\phi f) \int d^4 x \sqrt{-g}  \, \left[ a\, \dot \phi_0  \left[K^0\right]^{(4)} +  \delta \phi' \, \left[K^0\right]^{(3)} + \left(\partial_i \delta \phi\right) \, \left[K^i\right]^{(3)} \right] \, .
\end{align}
Schematically we have
\begin{equation}
K^\mu \sim \Gamma \, \partial \Gamma + \Gamma \,  \Gamma \, \Gamma \, . 
\end{equation}
The contributions to $\Gamma$ come through the constraints $N$ and $\tilde N_i$ as computed in GR. Differently from the computation of the cubic Lagrangian, here we need the solution for the constraints up to the second order in the perturbations. As we can see from Eq. \eqref{eq:constraints}, in spatially flat gauge the first order values for the constraints are suppressed by the slow-roll parameter as $\sqrt\epsilon$, while the second order values contain terms $\propto \delta\phi^2$ which do not possess any slow-roll suppression (see \cite{Seery:2006vu} for the computation). In detail, the auxiliary fields $N$ and $\tilde N_i$ are expanded up to second order as
\begin{align}
N & = 1 + \alpha \, , \\
\tilde N_j & = \partial_j \tilde\theta + \tilde\beta_j \, , \qquad \partial_j \tilde\beta^j = 0 \, .
\end{align}
Each one of these fields can be expanded in power of series of $\delta \phi$. Formally, we get 
\begin{align}
\alpha &= \alpha_1 + \alpha_2 + ... \, , \\
\tilde\theta &= \tilde\theta_1 + \tilde\theta_2 + ... \, , \\
\tilde\beta_j &= \tilde\beta_{1j} + \tilde\beta_{2j} + ... \, ,
\end{align}
where each index specifies the corresponding order in perturbation theory. 

The solutions to the constraint equations have been computed in \cite{Seery:2006vu} as\footnote{These results are converted in conformal time and the fields contributing to $\tilde N_j$ are rescaled by a factor $1/a$ with respect to the original reference.}
\begin{align}
\alpha_1 &= \sqrt{\frac{\epsilon}{2}} \, \delta \phi \, , \\
\tilde\theta_1 &= \partial^{-2}\left[- \sqrt{\frac{\epsilon}{2}} \, \delta\phi' \right] \\
& \qquad  + \partial^{-2}\left[a H \left(\epsilon - \eta\right) \sqrt{\frac{\epsilon}{2}} \, \delta \phi + a H \, \epsilon  \sqrt{\frac{\epsilon}{2}} \, \delta \phi \right] \, , \nonumber \\
\tilde\beta_{1j} &= 0 \, , \\
\alpha_2 &= \frac{1}{2 H} \partial^{-2}\Sigma \nonumber \\          
& \qquad + \frac{\alpha_1^2}{2} + \frac{1}{2 H} \partial^{-2}\left[\frac{1}{a^2} (\partial^{2} \alpha_1) (\partial^{2} \tilde\theta_1) - \frac{1}{a^2} (\partial_i \partial_j \alpha_1) (\partial_i \partial_j \tilde\theta_1) \right] \, , \\
\tilde\theta_2 &= \partial^{-2}\Big[- \frac{1}{4 H a} (\partial_i \delta \phi) (\partial_i \delta \phi) - \frac{1}{4 H a} \left(\delta \phi'\right)^2 - 3 a^2 H \, \alpha_2 \nonumber  \\
& + \frac{3}{2} a H \alpha_1^2 \left(3 - \epsilon \right) + a H \epsilon \, \alpha_2 - a \frac{\partial^2_{\phi \phi} V}{4 H} \left(\delta \phi\right)^2 + \frac{(\partial^2 \tilde\theta_1)^2}{4 H a^3} - \frac{(\partial_i\partial_j \tilde\theta_1 )^2}{4 H a^3} + \frac{1}{a} \sqrt{\frac{\epsilon}{2}} (\partial_i \tilde\theta_1) (\partial_i \delta \phi) \nonumber \\
& + 2 \frac{\alpha_1}{a} \left(\partial^2 \tilde\theta_1 + a \sqrt{\frac{\epsilon}{2}} \delta \phi' \right) \Big]\, , \label{eq:theta2} \\ 
\tilde\beta_{2j} &= 2 \, \partial^{-4}\left[(\partial_m \partial_j \delta \phi') (\partial_m \delta \phi) + (\partial_j \delta \phi') (\partial^2 \delta \phi) - (\partial^2 \delta \phi') (\partial_j\delta \phi) - (\partial_m \delta \phi') (\partial_m \partial_j \delta \phi) \right] + ... \nonumber\\
&=\tilde \beta^{even}_{2j}\, , \label{eq:beta_GR}
\end{align} 
where
\begin{equation} \label{eq:Sigma}
\Sigma = \frac{1}{a} \left[(\partial_j \delta\phi')\,(\partial_j \delta \phi) + \delta\phi' \, (\partial^2\delta \phi) \right] \, .
\end{equation}
The first lines in the previous equations represent the terms leading in slow-roll.

Now, we have all the ingredients to compute $\Delta S^{(4)}_{\rm CS}$. Using the slow-roll scaling of the auxiliary fields contributing to $\Gamma$ and the fact that $\Gamma^{(0)} = 0$, the contributions at leading order in slow-roll (those with less powers of the first order values of the auxiliary fields) come schematically from the terms
\begin{equation}
\mathcal L^{(4)} \sim a (\partial_\phi f) \,\dot \phi_0 \left(\Gamma^{(2)} \, \partial \Gamma^{(2)}\right) + (\partial_\phi f) \left[  \, \delta \phi' + \left(\partial_i \delta \phi\right)\right]   \left[\left(\Gamma^{(1)} \, \partial \Gamma^{(2)}\right) + \left(\Gamma^{(2)} \, \partial \Gamma^{(1)}\right)\right] \, , 
\end{equation}
where 
\begin{align}
{\Gamma^{(1) \mu}_{\alpha\beta}} = &\frac{\left[g^{\mu\lambda}\right]^{(0)}}{2} \left[\left(\partial_\alpha g_{\lambda\beta} + \partial_\beta g_{\alpha\lambda}  - \partial_\lambda g_{\alpha\beta}\right)\right]^{(1)}   \, , 
\end{align}
and 
\begin{align}
{\Gamma^{(2) \mu}_{\alpha\beta}} = \frac{\left[g^{\mu\lambda}\right]^{(0)}}{2} \left[\left(\partial_\alpha g_{\lambda\beta} + \partial_\beta g_{\alpha\lambda}  - \partial_\lambda g_{\alpha\beta}\right)\right]^{(2)} \, .
\end{align}
The metric components we need read
\begin{align} \label{eq:metric_quartic_int_pert}
 & g^{(1)}_{00} =  - 2 N^{(1)}  \, , \qquad g^{(1)}_{0i} = g^{(1)}_{i0} =  \tilde N^{(1)}_i \, , \qquad  g^{(1)}_{ij} = 0 \, , \nonumber \\
 & g^{(2)}_{00} =  - 2 N^{(2)}  \, , \qquad g^{(2)}_{0i} = g^{(2)}_{i0} =  \tilde N^{(2)}_i \, , \qquad  g^{(2)}_{ij} = 0 \, , \nonumber \\
 & \left[g^{00}\right]^{(0)} = - 1  \, , \qquad \left[g^{0i}\right]^{(0)} = \left[g^{i0}\right]^{(0)} =  0  \, , \qquad  \left[g^{ij}\right]^{(0)} = \delta_{ij} \, . 
\end{align}
From these, we derive the following values for the Christoffel symbols we are interested in
\begin{align}
\left[\Gamma^0_{00}\right]^{(1)} &= N'^{(1)}  \, , \qquad \qquad\qquad \qquad\qquad   \left[\Gamma^0_{00}\right]^{(2)} = N'^{(2)} \, , \\
\left[\Gamma^0_{0i}\right]^{(1)} &= \partial_i N^{(1)} \, , \qquad \qquad \qquad \qquad \quad \left[\Gamma^0_{0i}\right]^{(2)} = \partial_i N^{(2)} \, , \\
\left[\Gamma^{0}_{ij}\right]^{(1)} &= -\frac{1}{2} \partial_i \tilde N^{(1)}_j - \frac{1}{2}\partial_j \tilde N^{(1)}_i  \, ,  \qquad \quad  \left[\Gamma^{0}_{ij}\right]^{(2)} = -\frac{1}{2} \partial_i \tilde N^{(2)}_j - \frac{1}{2}\partial_j \tilde N^{(2)}_i  \, , \\
\left[\Gamma^i_{00}\right]^{(1)} &= \partial_i N^{(1)} + {\tilde N'^{(1)}_i} \, , \qquad \qquad \qquad  \left[\Gamma^i_{00}\right]^{(2)} = \partial_i N^{(2)} + {\tilde N'^{(2)}_i} \, ,   \\
\left[\Gamma^i_{0j}\right]^{(1)} &=  \frac{1}{2} \partial_j \tilde N^{(1)}_i - \frac{1}{2} \partial_i \tilde N^{(1)}_j \, , \qquad \qquad \left[\Gamma^i_{0j}\right]^{(2)} =  \frac{1}{2} \partial_j \tilde N^{(2)}_i - \frac{1}{2} \partial_i \tilde N^{(2)}_j \, , \\
\left[\Gamma^i_{jk}\right]^{(1)} &= 0 \, , \qquad \qquad \qquad \qquad \qquad \qquad \left[\Gamma^i_{jk}\right]^{(2)} = 0 \, .
\end{align}
Now, we proceed to compute the gCS current components as
\begin{align} 
\left[K^0\right]^{(3)} &= \epsilon^{0ijk}\left(\Gamma^\sigma_{i\rho}\partial_j\Gamma^\rho_{k\sigma}\right)  \nonumber \\ 
&= \epsilon^{ijk}\Bigg\{ \frac{1}{2} \left(\partial_i \partial_l \tilde \theta_1 \right) \left(\partial_j \partial_l \beta_{2k} \right) \Bigg\} \, , \label{eq:kappa0_current_4int} \\
\left[K^i\right]^{(3)} &= \epsilon^{ij0k}\left(\Gamma^\sigma_{j\rho}\partial_0\Gamma^\rho_{k\sigma}\right) + \epsilon^{i0jk}\left(\Gamma^\sigma_{0\rho}\partial_j\Gamma^\rho_{k\sigma}\right) + \epsilon^{ijk0}\left(\Gamma^\sigma_{j\rho}\partial_k\Gamma^\rho_{0\sigma}\right) \nonumber \\
&= \epsilon^{ijk} \Bigg\{\left(\partial_j \alpha_1 \right)\left(\partial_k \alpha'_2 \right) + \left(\partial_j \alpha_2 \right)\left(\partial_k \alpha'_1 \right) + \frac{1}{2} \left(\partial_j\partial_l \tilde\theta_1 \right) \left(\partial_l \tilde \beta'_{2k} \right) - \frac{1}{2} \left(\partial_j\partial_l \tilde \theta_1 \right) \left(\partial_k \tilde \beta'_{2l}  \right) \nonumber \\
& \qquad\qquad - \frac{1}{2} \left(\partial_j \tilde \beta_{2l}\right) \left(\partial_k \partial_l \tilde \theta_1' \right) + \frac{1}{2} \left(\partial_l \tilde \beta_{2j} \right) \left(\partial_k \partial_l \tilde \theta_1' \right) + 2 \left(\partial_l \alpha_1 \right) \left(\partial_l\partial_j \tilde \beta_{2k} \right) \nonumber \\
& \qquad\qquad + \left(\partial_l \tilde \theta_1' \right) \left(\partial_l\partial_j \tilde \beta_{2k} \right) \Bigg\} \, ,\label{eq:kappai_current_4int} \\
\left[K^0\right]^{(4)} &= \epsilon^{0ijk}\left(\Gamma^\sigma_{i\rho}\partial_j\Gamma^\rho_{k\sigma}\right)  \nonumber \\
&= \epsilon^{ijk}\Bigg\{\frac{1}{2} \left(\partial_l \partial_i \tilde \theta_2 \right) \left(\partial_l\partial_j \tilde \beta_{2k} \right) + \frac{1}{2} \left(\partial_l \tilde \beta_{2i }\right) \left(\partial_l\partial_j \tilde \beta_{2k} \right)  \Bigg\} \, .\label{eq:kappa0_current_4int_2x2}
\end{align}
Up to the leading-order in the slow-roll parameter, we have the following expressions,
\begin{align}
\alpha_1&\simeq \sqrt{\frac{\epsilon}{2}}\delta\phi,\\
\tilde\theta_1&\simeq - \sqrt{\frac{\epsilon}{2}} \partial^{-2}\delta\phi',\\
\tilde\beta_{1j}&=0,\\
\alpha_2&\simeq \frac{1}{2H}\partial^{-2}\Sigma_{\delta\phi},\\
\tilde\theta_2&\simeq-\frac{1}{2}\partial^{-2}\biggl[\frac{1}{2aH}(\partial_i\zeta)^2+\frac{1}{2aH}\zeta'^2+3a^2\partial^{-2}\Sigma_{\delta\phi}\biggr],\\
\tilde\beta_{2j}&\simeq 2\partial^{-4}[(\partial_m\partial_j\delta\phi')(\partial_m\delta\phi)+(\partial_j\delta\phi')(\partial^2\delta\phi)-(\partial^2\delta\phi')(\partial_j\delta\phi)-(\partial_m\delta\phi')(\partial_j\partial_m\delta\phi)] \, ,
\end{align}
where
\begin{align}
\Sigma_{\delta\phi}:=\frac{1}{a}[(\partial_j\delta\phi')(\partial_j\delta\phi)+\delta\phi'(\partial^2\delta\phi)] \, .
\end{align}
The quartic Lagrangian can be expressed in terms of the slow-roll parameter $\epsilon$ as
\begin{align}  \label{eq:intermediate:S_4}
\Delta S_{CS} = - 2 \sqrt{2\epsilon} \, \partial_\phi f(\phi_0) \int d^4 x  \left[ a H \left[K^0\right]^{(4)} + \frac{1}{\sqrt{2\epsilon}}\delta\phi' \left[K^0\right]^{(3)} + \frac{1}{\sqrt{2\epsilon}} \left(\partial_i \delta\phi\right) \left[K^i\right]^{(3)} \right] \, ,
\end{align}
where the Levi-Civita tensor $\epsilon^{ijk}$ is replaced by the Levi-Civita symbol  $\bar \epsilon^{ijk}$ in the $K^\mu$ components. 

Due to the properties of the Levi-Civita symbol, we can remove any term of the type $\partial_i(...) \partial_j(...)$ (or similar) in $\left[K^0\right]^{(4)}$ and $\left[K^i\right]^{(3)}$. Also, in general we can remove any term of the type $\partial_i\partial_j(...)$ (or similar) and exchange the indexes $i$, $j$ and $k$ with a switch of sign in the corresponding term. In the end we remain with the following expressions for each term in Eq. \eqref{eq:intermediate:S_4} 
\begin{align} 
a H \left[K^0\right]^{(4)} &= \bar\epsilon^{ijk}\Bigg\{\frac{a H}{2} \left(\partial_l \tilde \beta_{2i }\right) \left(\partial_l\partial_j \tilde \beta_{2k} \right)  \Bigg\} \, ,\\
\frac{1}{\sqrt{2\epsilon}}\delta\phi' \left[K^0\right]^{(3)}&= \frac{1}{2\sqrt{2\epsilon}} \bar\epsilon^{ijk}\Bigg\{ \delta\phi'\left(\partial_i \partial_l \tilde \theta_1 \right) \left(\partial_j \partial_l \tilde \beta_{2k} \right) \Bigg\} \, ,\\
\frac{1}{\sqrt{2\epsilon}}\left(\partial_i \delta\phi\right)  \left[K^i\right]^{(3)}&= \frac{1}{\sqrt{2\epsilon}}\bar\epsilon^{ijk} \Bigg\{\frac{1}{2} \left(\partial_i \delta\phi\right) \left(\partial_j\partial_l \tilde\theta_1 \right) \left(\partial_l \tilde \beta'_{2k} \right) + \frac{1}{2} \left(\partial_i \delta\phi\right) \left(\partial_l \tilde \beta_{2j} \right) \left(\partial_k \partial_l \tilde \theta_1' \right) \nonumber \\
& \qquad\qquad   + 2 \left(\partial_i \delta\phi\right) \left(\partial_l \alpha_1 \right) \left(\partial_l\partial_j \tilde \beta_{2k} \right) + \left(\partial_i \delta\phi\right) \left(\partial_l \tilde \theta_1' \right) \left(\partial_l\partial_j \tilde \beta_{2k} \right) \Bigg\} \, .
\end{align}
By doing appropriate integration by parts, the resulting Lagrangian density reads
\begin{align}  
\Delta \mathcal L^{(4,s)}_{CS} 
= - \frac{1}{4 M_{CS} H} \,\bar\epsilon^{ijk}\left[ \frac{a H}{2} \left(\partial_l \tilde \beta_{2i }\right) \left(\partial_l\partial_j \tilde \beta_{2k} \right)  +  \frac{2}{\sqrt{2\epsilon}} \left(\partial_i \delta\phi\right) \left(\partial_l \tilde \theta_1' + \partial_l \alpha_1  \right) \left(\partial_l\partial_j \tilde \beta_{2k} \right)  \right]\, .
\end{align}
Using the zero-th order equations of motion for $\delta\phi$, we have 
\begin{align} 
\theta_1' + \alpha_1 = -\sqrt{\frac{\epsilon}{2}}(\partial^{-2}\delta\phi'' - \delta\phi) =  2 a H \, \sqrt{\frac{\epsilon}{2}} (\partial^{-2}\delta\phi') \, ,
\end{align}
and the previous Lagrangian simplifies into 
\begin{align}   \label{eq:2orderCS}
\Delta \mathcal L^{(4,s)}_{CS,1} = - \frac{a}{4 M_{CS}} \,\bar\epsilon^{ijk}\left[ \frac{1}{2} \left(\partial_l \tilde \beta_{2i }\right) \left(\partial_l\partial_j \tilde \beta_{2k} \right)  + 2 \left(\partial_i \delta\phi\right) (\partial_l\partial^{-2}\delta\phi') \left(\partial_l\partial_j \tilde \beta_{2k} \right)  \right]\, ,
\end{align}
which is the gCS contribution to the parity-odd 4-scalars interactions.

Now, we move to the computation of the parity-odd contribution from the GR action, $S^{(4,s)}_{CS,2}$. The scalar quartic action in GR has been computed as~\cite{Seery:2006vu}
\begin{equation} \label{eq:GR_quartic_L}
S^{flat}_{GR} = \int d^4 x \, \left[ -\frac{1}{4} \beta_{2j} \left( \partial^2 \beta_{2j}\right) + a^2 \, \theta_2 \Sigma + \frac{3}{4} a^4 \, \left(\partial^{-2}\Sigma\right) \left(\partial^{-2}\Sigma\right) - a \, \delta \phi' (\partial_j \delta \phi) \beta_{2j} \right] \, ,
\end{equation}
where the untilded quantities (like $\beta_{2j}$) are obtained multipling by a factor of $a$ the tilded quantities (e.g. $\beta_{2j} = a \,\tilde \beta_{2j}$). Using the solutions for the constraints as above, this action provides parity-even scalar quartic interactions. However, the CS-term provides a quadratic correction to $\beta_{2j}$(or $\tilde\beta_{2j}$) that when plugged in Eq. \eqref{eq:GR_quartic_L} provides a new set of scalar parity-odd quartic interactions. Conversely, scalar fields such as $\alpha_2$ and $\theta_2$ cannot contain parity-odd quadratic corrections in terms of $\delta\phi$ in $S_{\rm slow-roll}$ as they must appear in combinations like
\begin{equation}
\propto \bar\epsilon^{ijk} (\partial_i\delta\phi) (\partial_j\partial_k\delta\phi)
\end{equation}
and similar combinations taking time-derivatives of $\delta\phi$. These are all zero due to the properties of the Levi-Civita symbol. 

We thus decompose $\beta_{2j}(\tilde\beta_{2j})$ into parity-odd and parity-even parts:
\begin{align}
\beta_{2j}=\beta^{even}_{2j}+\beta^{odd}_{2j}  \, , \qquad \mbox{or} \qquad \tilde\beta_{2j}=\tilde\beta^{even}_{2j}+\tilde\beta^{odd}_{2j} \, ,
\end{align}
where $\tilde\beta^{even}_{2j}$ is as in Eq. \eqref{eq:beta_GR}. Here, we derive $\beta^{odd}_{2j}$ for the first time in CS gravity. The correction to the equation of motion for $\beta_{2i}$ at leading order in slow-roll can be read once applying the functional derivative with respect to $\beta^s_{2}$ to Eq. \eqref{eq:2orderCS} once expressed it in terms of the untilded $\beta^i_{2}$'s
\begin{align}   
\Delta \mathcal L_{CS} = - \frac{1}{4 a M_{CS}} \,\bar\epsilon^{ijk}\left[ \frac{1}{2} \left(\partial_l \beta^i_{2 }\right) \left(\partial_l\partial_j \beta^k_{2} \right)  + 2 a \left(\partial_i \delta\phi\right) (\partial_l\partial^{-2}\delta\phi') \left(\partial_l\partial_j \beta^k_{2} \right)  \right]\, .
\end{align}
We then get
\begin{align}   
\frac{\delta\Delta \mathcal L_{CS}}{\delta\beta^s_{2}} = - \frac{1}{4 a M_{CS}} \,\left[ \bar\epsilon^{ijs} \left(\partial^2\partial_j  \beta_{2i} \right) + 2 a \bar\epsilon^{ijs} \left(\partial_i \partial_l\delta\phi\right) (\partial_l\partial^{-2}\partial_j\delta\phi') + 2 a \bar\epsilon^{ijs} \left(\partial_i \delta\phi\right) (\partial_j\delta\phi')  \right]\, .
\end{align}
The resulting equation of motion once adding the contribution from standard GR to the functional derivative reads
\begin{align}   
&-\frac{1}{2} \partial^2\beta_{2s} - \frac{1}{4 a M_{CS} } \,\left[ \bar\epsilon^{ijs} \left(\partial^2\partial_j  \beta_{2i} \right) + 2 a \bar\epsilon^{ijs} \left(\partial_i \partial_l\delta\phi\right) (\partial_l\partial^{-2}\partial_j\delta\phi') + 2 a \bar\epsilon^{ijs} \left(\partial_i \delta\phi\right) (\partial_j\delta\phi')  \right] \nonumber \\
&\ = (\mbox{GR parity-even terms}) \, . 
\end{align}
From this equation we can derive the parity-odd part of $\beta_{2j}$ at leading order in slow-roll:
\begin{align}  \label{eq: beta-odd}
\beta^{odd}_{2s} = - \frac{1}{2a M_{CS}} \,\bar\epsilon^{ijs}\partial^{-2}\left[  \left(\partial^2\partial_j  \beta^{even}_{2i} \right) + 2 a  \left(\partial_i \partial_l\delta\phi\right) (\partial_l\partial^{-2}\partial_j\delta\phi') + 2 a  \left(\partial_i \delta\phi\right) (\partial_j\delta\phi')  \right]  \, . 
\end{align}
By substituting Eq.~(\ref{eq: beta-odd}) back in Eq. \eqref{eq:GR_quartic_L}, we obtain
 \begin{align}   \label{eq:newGR_odd}
\Delta \mathcal L^{(4,s)}_{CS,2} &= \left[ -\frac{1}{4} \beta^{odd}_{2k} \left( \partial^2 \beta^{even}_{2k}\right) -\frac{1}{4} \beta^{even}_{2k} \left( \partial^2 \beta^{odd}_{2k}\right) - a \, \delta \phi' (\partial_k \delta \phi) \beta^{odd}_{2k} \right] \nonumber \\ 
& = \frac{1}{2a M_{CS}} \,\bar\epsilon^{ijk} \Big[ \frac{1}{4} \left(\partial_j  \beta^{even}_{2i} \right) \left( \partial^2 \beta^{even}_{2k}\right) + \frac{1}{2} a \,\partial^{-2}\left[(\partial_l\partial^{-2}\partial_j\delta\phi')\left(\partial_i \partial_l\delta\phi\right) \right] \left( \partial^2 \beta^{even}_{2k}\right)  \nonumber \\ 
& \qquad\qquad\qquad\ \ + \frac{1}{2} a  \,\partial^{-2}\left[(\partial_j\delta\phi')\left(\partial_i \delta\phi\right) \right] \left( \partial^2 \beta^{even}_{2k}\right)+\frac{1}{4} \left(\partial^2\partial_j  \beta^{even}_{2i} \right) \beta^{even}_{2k}  \nonumber \\
& \qquad\qquad\qquad\ \ + \frac{1}{2} a \,(\partial_l\partial^{-2}\partial_j\delta\phi') \left(\partial_i \partial_l\delta\phi\right)  \beta^{even}_{2k} + \frac{1}{2} a \,  (\partial_j\delta\phi')  \left(\partial_i \delta\phi\right)\beta^{even}_{2k} \nonumber \\
& \qquad\qquad\qquad\ \ +a \,\delta \phi' (\partial_k \delta \phi)\left(\partial_j  \beta^{even}_{2i} \right)  + 2 a^2 \, \partial^{-2}\left[(\partial_l\partial^{-2}\partial_j\delta\phi') \left(\partial_i \partial_l\delta\phi\right) \right]  \delta \phi' (\partial_k \delta \phi) \nonumber \\
& \qquad\qquad\qquad\ \ + 2 a^2 \, \partial^{-2}\left[(\partial_j\delta\phi') \left(\partial_i\delta\phi\right) \right]  \delta \phi' (\partial_k \delta \phi) \Big] \, ,
\end{align}
which are the additional quartic interaction terms due to the modification of the second order constraint $\beta_{2j}$ provided by the gCS term. The scalar quartic Lagrangian from the gCS term at leading order in slow-roll has been computed for the first time in the present paper.

\section{Scalar trispectrum}
\label{sec:5}
In the following subsections, we study both graviton-mediated and contact diagrams. So far, parity-violating signatures in a scalar trispectrum from a graviton-mediated diagram have been explored in the context of the dynamical Chern-Simons gravity~\cite{Creque-Sarbinowski:2023wmb} in which only GR vertexes were considered (see also the more recent work \cite{Stefanyszyn:2025yhq}). To obtain parity violation, the authors of~\cite{Creque-Sarbinowski:2023wmb} have taken into account the parity violation in the tensor mode function. In our paper, we consider parity-violation coming from Chern-Simons interaction vertices, while the tensor mode function is the one of GR, Eq. \eqref{eq:mode_function}.

\subsection{Graviton-mediated trispectrum}

\begin{figure}
        \centering
        \includegraphics[width=14cm]{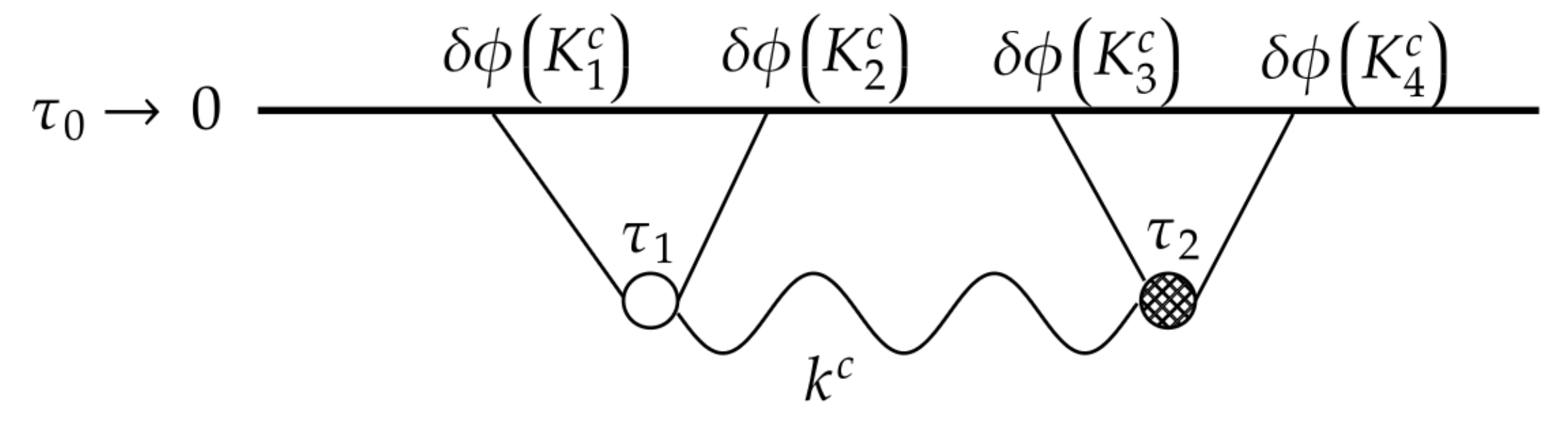} 

    \caption{Diagram of the scalar parity-odd trispectrum mediated by graviton exchange. The external legs represent scalar propagators, while the wavy line represents the graviton propagator. The thick horizontal line is the conformal boundary $\tau_0 \rightarrow 0$, so that the graviton propagator lives in the bulk, $\tau_i <\tau_0$. The white vertex comes from the standard kinetic term of the inflaton in Eq. \eqref{eq:Lsst_GR}. The filled vertex corresponds to the Chern-Simons coupling in Eq. \eqref{eq:sst_CS_new}. 
    } \label{fig:CS_F} 
\end{figure}

By using the Schwinger-Keldysh (SK) diagrammatic formalism outlined in Ref. \cite{Chen:2017ryl}, the first leading contribution to the parity-odd trispectrum arising from the gCS term originates from the graviton-exchange diagram depicted in Fig. \ref{fig:CS_F}. In this diagram, one vertex corresponds to the standard GR scalar-scalar-tensor interaction, while the other involves the scalar-scalar-tensor vertex induced by the gCS term. The necessity of only a single gCS vertex in this context is a consequence of the diagram’s behavior under parity transformations: GR propagators and vertices remain invariant under parity, whereas the gCS-induced vertex changes sign. Consequently, parity violation manifests only in diagrams containing an odd number of gCS vertices. In contrast, diagrams involving an even number of gCS vertices are parity-even and therefore do not contribute to the parity-odd component of the trispectrum.

The GR-vertex is already computed in the literature (see e.g. \cite{Maldacena:2002vr}) and reads
\begin{align} \label{eq:Lsst_GR}
\left[\Delta \mathcal L_{GR}\right]^{(3)sst, \, flat} =   \frac{1}{2}\, a^2 \left[ (\partial_i \delta\phi) \left(\partial_j \delta\phi \right) h_{ij} \right] \, .
\end{align}
Converting Eq. \eqref{eq:Lsst_GR} in Fourier space and using $a \simeq - (H \tau)^{-1}$, we find
\begin{align} \label{eq:Lsst_GR_fourier}
\left[\Delta \mathcal L_{GR}\right]^{(3)sst, \, flat} =  - \frac{1}{2 H^2 \tau^2} \, \delta\phi(\mathbf{p_1}) \, \delta\phi(\mathbf{p_2}) \,h_{\lambda}(\mathbf{p_3}) \left(p^i_1 p^j_2 \cdot e^{\lambda}_{ij}(\hat{p_3})\right) \, .
\end{align}
The full computation of the scalar-scalar-tensor cubic interactions introduced by the gCS-term is given above in Sec.~\ref{Appendix: perturbed actions_sst} and the final result of the parity-odd cubic Lagrangian is given in Eq. \eqref{eq:sst_CS_new}. Converting Eq. \eqref{eq:sst_CS_new} in Fourier space and using $a \simeq - (H \tau)^{-1}$, we find
\begin{align} \label{eq:Lsst_com_CS2}
\left[\Delta \mathcal L_{CS}\right]^{(3)sst, \, flat} = -  \frac{1}{2 H M_{CS}}\, \frac{1}{\tau} \, \alpha_\lambda \, \frac{p_3}{p_2^2} \, \delta\phi(\mathbf{p_1}) \, \delta\phi'(\mathbf{p_2}) \,h'_{\lambda}(\mathbf{p_3}) \left(p^i_1 p^l_2 \cdot e^{\lambda}_{il}(\hat{p_3})\right)\, ,
\end{align}
where the sum over the polarization index $\lambda$ is understood.

Following the prescriptions of the SK rules, we compute the graviton-mediated contribution via these steps:
\begin{enumerate}
\item We label the external legs with the momenta $\bk_i$ and consider all the possible independent combinations. Since the two vertexes are different and the order of the $\delta\phi$ fields is important in the filled vertex, we have 12 such combinations, and each of these combinations is a \textit{channel} over which we have to sum. We define each of these 12 combinations with an ordered set of momenta $\mathbf{K^c}$. In the following, we have the 12 independent sets:
\begin{align}
\mathbf{K^{s_1}} = \{ \bk_1,\bk_2,\bk_3,\bk_4 \} \, , \qquad\qquad \mathbf{K^{t_1}} = \{ \bk_1,\bk_3,\bk_2,\bk_4 \} \, , \qquad\qquad \mathbf{K^{u_1}} = \{ \bk_1,\bk_4,\bk_2,\bk_3 \} \nonumber \, , \\ 
\mathbf{K^{s_2}} = \{ \bk_1,\bk_2,\bk_4,\bk_3 \} \, , \qquad\qquad \mathbf{K^{t_2}} = \{ \bk_1,\bk_3,\bk_4,\bk_2 \} \, , \qquad\qquad \mathbf{K^{u_2}} = \{ \bk_1,\bk_4,\bk_3,\bk_2 \}  \, , \nonumber \\ 
\mathbf{K^{s_3}} = \{ \bk_3,\bk_4,\bk_1,\bk_2 \} \, , \qquad\qquad \mathbf{K^{t_3}} = \{ \bk_2,\bk_4,\bk_1,\bk_3 \} \, , \qquad\qquad \mathbf{K^{u_3}} = \{ \bk_2,\bk_3,\bk_1,\bk_4 \}  \nonumber \, , \\ 
\mathbf{K^{s_4}} = \{ \bk_3,\bk_4,\bk_2,\bk_1 \} \, , \qquad\qquad \mathbf{K^{t_4}} = \{ \bk_2,\bk_4,\bk_3,\bk_1 \} \, , \qquad\qquad \mathbf{K^{u_4}} = \{ \bk_2,\bk_3,\bk_4,\bk_1 \} \nonumber \, . \\  
\end{align}
The Latin letter at apex labels the Mandelstam variable associated to each combination, which is labeled 
\begin{align}
\mathbf{k^{c}} = \mathbf{K_1^{c}}+ \mathbf{K_2^{c}} = -\mathbf{K_3^{c}} - \mathbf{K_4^{c}} \, .
\end{align}
For example if $c=s_1$, then $\mathbf{k^{s_1}} = \bk_1+ \bk_2 = -\bk_3-\bk_4$. Moreover, we include a symmetry factor $2$ due to the symmetry of the white vertex in the exchange of the $\mathbf{K_1^{c}}$ and $\mathbf{K_2^{c}}$ momenta. 
\item For each channel, we can label each vertex with either a $+$ and $-$. Therefore, in total there will be four combinations, $++$, $+-$, $-+$, and $--$, for every channel. We integrate over the time $\tau_1$ in the white vertex and over the time $\tau_2$ in the filled vertex. Each time integration goes from $-\infty$ to $\tau_0$, where $\tau_0$ here denotes the time at the end of inflation ($\tau_0 \rightarrow 0$). For each $\pm$ vertex multiply for $\mp i$, respectively. For the white vertex, we add the corresponding coupling in Eq. \eqref{eq:Lsst_GR_fourier} with the opposite sign: $1/(2 H^2 \tau^2)$. For the filled vertex, we add the corresponding coupling in Eq. \eqref{eq:Lsst_com_CS2} with the opposite sign: $\alpha_\lambda/(2H M_{CS}\tau)$. 
\item Each external leg connected to the vertex corresponds to one factor of the following scalar bulk-to-boundary propagator (within the corresponding time integral):
\begin{align}
\mathcal{G}^{\rm no \ deriv}_{\pm}(\tau,k) &= \frac{H^2}{2 k^3} \, (1 \mp i k \tau) \, e^{\pm i k \tau} \, e^{\mp i k \tau_0} \, (1 \pm i k \tau_0)  \, , \nonumber \\
\end{align}
or 
\begin{align}
\mathcal{G}^{\rm deriv}_{\pm}(\tau,k) &= \frac{H^2}{2 k} \, \tau \, e^{\pm i k \tau} \, \, e^{\mp i k \tau_0} \, (1 \pm i k \tau_0) \, . 
\end{align}
Here, the $\pm$ is related to the vertex which each external leg is connected to. Also, the ``no deriv" applies when the interaction leg is without time derivatives, while in the case in which there is a time derivative the ``deriv" case applies. The scalar propagator has momentum equal to the momentum of the boundary scalar perturbation of the external leg and is evaluated at the conformal time of the associated vertex. \textit{Boundary} refers to the conformal time boundary $\tau_0 \rightarrow 0$ and \textit{bulk} to $\tau_1, \tau_2 < \tau_0$. 
\item For each vertex combination, we need to multiply by the graviton-helicity-dependent (but time-independent) vertex coupling of the form
\begin{align}
\mathcal C^{CS}_\lambda(\mathbf{K^c}) =  \frac{k^c}{(K_4^c)^2} \, \left[e^{(\lambda)}_{ij}(\hat k^c) \, (K_1^c)^i \, (K_2^c)^j \right] \, \left[e^{(\lambda)*}_{ij}(\hat k^c) \, (K_3^c)^i \, (K_4^c)^j \right].
\end{align}
This polarization portion obeys to the relation $\mathcal C^{CS*}_R(\mathbf{K^c}) = \mathcal C^{CS}_L(\mathbf{K^c})$.  
\item Then, we write down one of the corresponding bulk propagators for a graviton of helicity~$\lambda$:
\begin{align}
&++ \qquad u_\lambda (\tau_1, k^c) \, u'^{*}_\lambda (\tau_2, k^c) \, \Theta(\tau_1 - \tau_2) + u^*_\lambda (\tau_1, k^c) \, u'_\lambda (\tau_2, k^c) \, \Theta(\tau_2 - \tau_1) \, , \\
&+- \qquad u^*_\lambda (\tau_1, k^c) \, u'_\lambda (\tau_2, k^c) \, , \\
&-+ \qquad u_\lambda (\tau_1, k^c) \, u'^{*}_\lambda (\tau_2, k^c) \, , \\
&-- \qquad  u^{*}_\lambda (\tau_1, k^c) \, u'_\lambda (\tau_2, k^c) \, \Theta(\tau_1 - \tau_2) + u_\lambda (\tau_1, k^c) \, u'^*_\lambda (\tau_2, k^c) \, \Theta(\tau_2 - \tau_1) \, . 
\end{align}
Here the mode-function $u_\lambda$ corresponds to the one in Eq. \eqref{eq:mode_function} and $u'_\lambda$ is its conformal time derivative. Also,  $\Theta(x)$ stands for the Heaviside function.
\item In the end for each term we need to sum over all the vertex combinations and channels to obtain the final result.
\end{enumerate}

Implementing these rules, the trispectrum for the graviton-mediated case reads\footnote{Here and afterwards the notation of the expectation value with the prime,  $\langle ... \rangle'$, indicates that we are implying the momentum conservation Dirac delta: $(2 \pi)^3 \delta^{(3)}(\sum_i {\bf k}_i)$.}
\begin{align}
\langle\delta\phi({\bf k}_1)\delta\phi({\bf k}_2)\delta\phi({\bf k}_3)\delta\phi({\bf k}_4)\rangle'=  \frac{1}{4M_{CS}H^3}\sum_{\lambda,c}\alpha_\lambda \, \mathcal{C}^{\rm CS}_\lambda({\bf K}^c){\rm Tvar}^{\rm CS}_\lambda({\bf K}^c) \, ,
\end{align}
where
\begin{align}
\mathcal{C}^{\rm CS}_\lambda({\bf K}^c)&:=\frac{k^c}{(K_4^c)^2}\left[e^{(\lambda)}_{ij}(\hat k^c)(K_1^c)^i(K_2^c)^j\right]\left[e^{(\lambda)*}_{kl}(\hat k^c)(K_3^c)^k(K_4^c)^l\right],\\
\label{time_ordered_graviton_med} {\rm Tvar}^{\rm CS}_\lambda({\bf K}^c)&:=-2{\rm Re}\left[\mathcal{I}^{(\lambda)}_1({\bf K}^c)+\mathcal{I}^{(\lambda)}_2({\bf K}^c)-\mathcal{I}^{(\lambda)}_3({\bf K}^c)\right]\, ,
\end{align} 
with
\begin{align}
\mathcal{I}^{(\lambda)}_1({\bf K}^c)&:=\int_{-\infty}^{\tau_0}\frac{d \tau_1}{\tau_1^2} \, \mathcal{G}_+^{\rm no\ deriv}(\tau_1,K_1^c) \, \mathcal{G}_+^{\rm no\ deriv}(\tau_1,K_2^c)\, u_\lambda(\tau_1,k^c)\notag\\
&\quad\quad \times\int_{-\infty}^{\tau_1}\frac{d\tau_2}{\tau_2} \, \mathcal{G}_+^{\rm no\ deriv}(\tau_2,K_3^c) \, \mathcal{G}_+^{\rm deriv}(\tau_2,K_4^c) \, u^{*'}_\lambda(\tau_2,k^c) \, ,\\
\mathcal{I}^{(\lambda)}_2({\bf K}^c)&:=\int_{-\infty}^{\tau_0}\frac{d \tau_2}{\tau_2} \, \mathcal{G}_+^{\rm no \ deriv}(\tau_2,K_3^c) \, \mathcal{G}_+^{\rm deriv}(\tau_2,K_4^c) \, u'_\lambda(\tau_2,k^c)\notag\\
&\quad\quad \times\int_{-\infty}^{\tau_2}\frac{d\tau_1}{\tau_1^2} \, \mathcal{G}_+^{\rm no\ deriv}(\tau_1,K_1^c) \, \mathcal{G}_+^{\rm no\ deriv}(\tau_1,K_2^c) \, u^{*}_\lambda(\tau_1,k^c) \, ,\\
\mathcal{I}^{(\lambda)}_3({\bf K}^c)&:=\int_{-\infty}^{\tau_0}\frac{d\tau_1}{\tau_1^2} \, \mathcal{G}_+^{\rm no\ deriv}(\tau_1,K_1^c) \, \mathcal{G}_+^{\rm no\ deriv}(\tau_1,K_2^c) \, u^{*}_\lambda(\tau_1,k^c) \nonumber \\
&\quad\quad \times \int_{-\infty}^{\tau_0}\frac{d\tau_2}{\tau_2} \, \mathcal{G}_-^{\rm no \ deriv}(\tau_2,K_3^c) \, \mathcal{G}_-^{\rm  deriv}(\tau_2,K_4^c) \, u'_\lambda(\tau_2,k^c) \, .
\end{align}
Up to constant coefficients (i.e., $\mathcal{I}^{(\lambda)}_i({\bf K}^c)=(\cdots) \times \bar{\mathcal{I}}^{(\lambda)}_i({\bf K}^c)$), these integrals read
\begin{align}
\bar{\mathcal{I}}^{(\lambda)}_1({\bf K}^c)&=(1+i K_1^c \tau_0) (1+i K_2^c \tau_0) (1+i K_3^c \tau_0) (1+i K_4^c \tau_0) \, e^{- i (K_1^c+K_2^c+K_3^c+K_4^c)\tau_0} \notag\\
&\quad\quad\times\int_{-\infty}^{\tau_0}\frac{d \tau_1}{\tau_1^2} (1-i K_1^c \tau_1)  (1-i K_2^c \tau_1) (1+i k^c \tau_1) \, e^{i (K_1^c+K_2^c- k^c)\tau_1} \notag\\
&\quad\quad \times\int_{-\infty}^{\tau_1} d\tau_2 \,\tau_2 \, (1-i K_3^c \tau_2) \, e^{i (K_3^c+K_4^c+k^c)\tau_2} \, , \label{eq:Ibar1} \\
\bar{\mathcal{I}}^{(\lambda)}_2({\bf K}^c)&=(1+i K_1^c \tau_0) (1+i K_2^c \tau_0) (1+i K_3^c \tau_0) (1+i K_4^c \tau_0) \, e^{- i (K_1^c+K_2^c+K_3^c+K_4^c)\tau_0} \notag\\ 
&\quad\quad \times \int_{-\infty}^{\tau_0} d \tau_2 \, \tau_2 \,  (1-i K_3^c \tau_2) \, e^{i (K_3^c+K_4^c-k^c)\tau_2} \notag\\
&\quad\quad \times \int_{-\infty}^{\tau_2} \frac{d\tau_1}{\tau_1^2} (1-i K_1^c \tau_1)  (1-i K_2^c \tau_1) (1-i k^c \tau_1) \, e^{i (K_1^c+K_2^c+k^c)\tau_1}  \, , \label{eq:Ibar2} \\
\bar{\mathcal{I}}^{(\lambda)}_3({\bf K}^c)&= (1+i K_1^c \tau_0) (1+i K_2^c \tau_0) (1-i K_3^c \tau_0) (1-i K_4^c \tau_0) \, e^{- i (K_1^c+K_2^c-K_3^c-K_4^c)\tau_0} \notag\\ 
&\quad\quad \times \int_{-\infty}^{\tau_0}\frac{d\tau_1}{\tau_1^2}  (1-i K_1^c \tau_1)  (1-i K_2^c \tau_1) (1-i k^c \tau_1) \, e^{i (K_1^c+K_2^c+k^c)\tau_1}  \notag\\
&\quad\quad \times \int_{-\infty}^{\tau_0} d\tau_2 \, \tau_2 \, (1+i K_3^c \tau_2)  \, e^{-i (K_3^c+K_4^c+k^c)\tau_2} \, . \label{eq:Ibar3} 
\end{align}
By performing these time integrations analytically using the usual i$\epsilon$-prescription at infinity, we find 
\begin{align}
{\rm Re}\left[\bar{\mathcal{I}}^{(\lambda)}_1({\bf K}^c)\right]&= -\frac{3 K_3^c + K_4^c + k^c}{(K_3^c+K_4^c+k^c)^3 \tau_0}\, ,\\
{\rm Re}\left[\bar{\mathcal{I}}^{(\lambda)}_2({\bf K}^c)\right]&= 0\, ,\\
{\rm Re}\left[\bar{\mathcal{I}}^{(\lambda)}_3({\bf K}^c)\right]&= -\frac{3 K_3^c + K_4^c + k^c}{(K_3^c+K_4^c+k^c)^3 \tau_0} \, .
\end{align}
It follows 
\begin{align}
{\rm Tvar}^{\rm CS}_\lambda({\bf K}^c) = 0 \, .
\end{align}
We find a vanishing result. This is consistent with the no-go theorem (see e.g. \cite{Cabass:2022rhr}) as there are no net infra-red (logarithmic) divergences in the in-in integrals in the limit $\tau_0 \rightarrow 0$.

\subsection{Contact-diagram trispectrum} \label{subsection:contact}

\begin{figure}
        \centering
        \includegraphics[width=14cm]{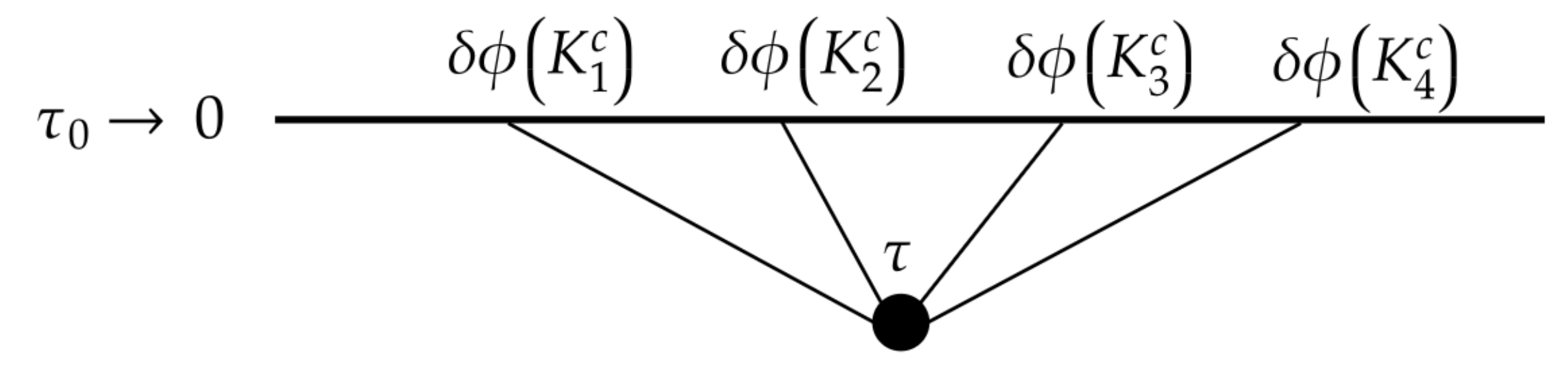} 

    \caption{Contact diagram contribution for the scalar parity-odd trispectrum. The black vertex can be each one of the interactions in Eqs. \eqref{eq:contactCS} and \eqref{eq:contactGR}.} \label{fig:CS_contact} 
\end{figure}

The second leading contribution to the parity-odd trispectrum comes from the contact diagram depicted in Fig. \ref{fig:CS_contact}. Here we need the scalar quartic interactions induced by the gCS term. These have been computed in Sec.~\ref{sec:4_quartic}: in total we find two independent contributions: one of these contributions come by substituting the GR-solutions of the auxiliary fields back to the gCS-Lagrangian ($CS,1$), Eq. \eqref{eq:2orderCS},  while another contribution comes by substituting the gCS parity-odd modification of the auxiliary field $\beta_i$ back to the GR-Lagrangian ($CS,2$), Eq. \eqref{eq:newGR_odd}. By substituting $a = - 1/H\tau$ and doing some algebra, the parity-odd quartic Lagrangian contributions in Fourier space read 
\begin{align}  \label{eq:contactCS}
\Delta \mathcal L^{(4,s)}_{CS,1}  =& i  \frac{1}{M_{CS} H \tau} \frac{1}{p^2_{12}} \left[ \frac{1}{2} + \frac{(\mathbf p_{1} \cdot \mathbf p_{12})}{p_1^2} \right] \, \Big[\mathbf p_1 \cdot (\mathbf p_{2} \times \mathbf p_3) \Big]  \delta\phi'(\mathbf p_1) \delta\phi(\mathbf p_2) \delta\phi'(\mathbf p_3) \delta\phi(\mathbf p_4) \, , 
\end{align}
and
 \begin{align}   \label{eq:contactGR}
\Delta \mathcal L^{(4,s)}_{CS,2} &= 
\frac{i}{2 M_{CS} H \tau} \, \Big[\left(\frac{1}{p^2_{12}}  - \frac{(\mathbf p_{1} \cdot \mathbf p_{2})}{p^2_{1} \, p^2_{12}}  - \frac{4}{p^2_{12}} +\frac{1}{p^2_{12}} - \frac{(\mathbf p_{1} \cdot \mathbf p_{2})}{p^2_{1}  \, p^2_{12}}\right)  \nonumber \\
& \quad\ + \frac{2}{p^2_{12}} \left(\frac{(\mathbf p_{1} \cdot \mathbf p_{2})}{p^2_{1}}+1\right)\Big] \Big[\mathbf p_1 \cdot (\mathbf p_{2} \times \mathbf p_3) \Big] \,  \delta\phi'(\mathbf p_1) \delta\phi(\mathbf p_2) \delta\phi'(\mathbf p_3) \delta\phi(\mathbf p_4) \nonumber \\
&= 0 \, , 
\end{align}
where we used $\sum_i{\bf p}_i={\bf 0}$ (momenta conservation), and $p_{12}=p_{34}$, with $p_{ij}=|{\bf p}_i+{\bf p}_j|$. Note that after we enforce momenta conservation at the 4-scalars vertex the interactions ($CS,2$) in Eq. \eqref{eq:contactGR} vanish, and therefore only interactions in Eq. \eqref{eq:contactCS} contribute to the contact-diagram. 

All the interactions have the same time-scaling, therefore apart for momenta-dependent contractions they are all proportional to the same in-in time integral. Using the S-K formalism as above, from the 4-legs contact diagram (Fig. \ref{fig:CS_contact}) we get the following contribution to the scalar trispectrum 
\begin{align}
\langle\delta\phi({\bf k}_1)\delta\phi({\bf k}_2)\delta\phi({\bf k}_3)\delta\phi({\bf k}_4)\rangle' =& -i  \frac{H^7}{16 M_{CS}} \frac{1}{k_{1} k^3_{2} k_{3} k^3_{4} k^2_{12}} \left[ \frac{1}{2} + \frac{(\mathbf k_{1} \cdot \mathbf k_{12})}{k_1^2} \right] \nonumber \\
&\qquad\qquad\times \Big[\mathbf k_1 \cdot (\mathbf k_{2} \times \mathbf k_3) \Big] \, I_{contact}(k_i)  +\mbox{perms}(k_i) \, ,
\end{align}
where 
\begin{align} \label{eq:time_integral_contact}
I_{contact}(k_i) & = 2\,{\rm Im}\left[\prod_i (1+i k_i \tau_0) \, e^{-i k_i \tau_0}\int_{-\infty}^{\tau_0} d\tau \, \tau \, (1-i k_2\tau)(1-i k_4\tau) \, e^{i (k_1+k_2+k_3+k_4)\tau}\right] \nonumber \\
& = 0 \quad (\tau_0 \rightarrow 0)\, ,
\end{align}
which indicates
\begin{align}
\langle\delta\phi({\bf k}_1)\delta\phi({\bf k}_2)\delta\phi({\bf k}_3)\delta\phi({\bf k}_4)\rangle'=0 \, .
\end{align}
We have thus shown that both contributions are found to vanish as we should expect due to the no-go theorem for parity-odd scalar trispectra in absence of (logarithmic) infra-red divergences in the in-in time-integrals. 

\section{Discussion: towards obtaining a non-vanishing parity-odd trispectrum} \label{sec:6}

So far, we have confirmed that the scalar parity-odd trispectrum from the zero-th order mode functions vanishes under the assumption that both scalar and tensor modes are in the Bunch-Davies initial state, which is one of the assumptions of the no-go theorem in Ref.~\cite{Cabass:2022rhr}. Thus, one may expect to obtain a nonvanishing parity-odd scalar trispectrum when introducing non-Bunch-Davies initial conditions. In this section, we verify that the above statement is correct. To do so, we first introduce the following simple modification to the scalar mode-function:
\begin{align} \label{eq:NBD_scalar}
\delta\phi^{\rm scalar}_k=\alpha_k\delta\phi^{\rm scalar}_{k,BD}+\beta_k\delta\phi^{\rm scalar *}_{k,BD} \, ,
\end{align} 
where $\delta\phi_{k,BD}$ is the Bunch-Davies scalar mode-function in Eq. \eqref{eq:scalar_BD}, and $\alpha_k$ and $\beta_k$ are the so-called Bogoliubov coefficients. Here we assume that these coefficients are constant complex numbers, i.e.\begin{align} \label{eq:bolig}
\alpha_k&:=1+iC_1 \, , \nonumber\\
\beta_k&:=C_2+iC_3 \, .
\end{align}
It has been known that the excited modes generally cause backreaction on the inflationary background and modification to the power spectra, which put constraints on the magnitude of $\alpha_k$ and $\beta_k$ (see, e.g., ~\cite{Tanaka:2000jw}). In order to avoid dealing with these issues, here we assume $C_i$'s to be small real constants satisfying $|C_i|\ll1$, which in fact prevent relevant backreaction to the background and power spectra corrections. As a result, the effect that we will compute, while non-zero, will remain unobservably small. Indeed, it would be interesting to explore concrete models in which the current observational constraints on primordial correlation functions are satisfied, while the parity-odd trispectrum is amplified up to an observable level. We will leave this to our future work, and here we only want to explicitly show that a non-zero result can be obtained via non-BD states.

Based on the above setup, let us study the impacts of non-Bunch-Davies states on the trispectrum from the graviton-exchange diagram. By imposing the non-Bunch-Davies conditions for the scalar modes, the scalar bulk-to-boundary propagator is modified. As a result, we obtain the following corrections to the integrals $\bar{\mathcal{I}}^{(\lambda)}_i({\bf K}^c)$ in Eqs. \eqref{eq:Ibar1}-\eqref{eq:Ibar3} 
\begin{align}
\Delta({\rm Re}[\bar{\mathcal{I}_1}({\bf K^c})])&=C_2\biggl[\frac{-3K_3^c+K_4^c-k^c}{(K_3^c-K_4^c+k^c)^3}+\frac{-3K_3^c+K_4^c+k^c}{(K_3^c-K_4^c-k^c)^3}-\frac{6(K_3^c+K_4^c+k^c)}{(K_3^c+K_4^c+k^c)^3}\biggr]\frac{1}{\tau_0}+C_3\mathcal{O}(1/k_i) \, ,\notag\\
\Delta(\rm Re[\bar{\mathcal{I}_2}({\bf K^c})])&=C_3\mathcal{O}(1/k_i) \, ,\notag\\
\Delta(\rm Re[\bar{\mathcal{I}_3}({\bf K^c})])
&=C_2\biggl[\frac{-3K_3^c+K_4^c-k^c}{(K_3^c-K_4^c+k^c)^3}+\frac{-3K_3^c+K_4^c+k^c}{(K_3^c-K_4^c-k^c)^3}-\frac{6(K_3^c+K_4^c+k^c)}{(K_3^c+K_4^c+k^c)^3}\biggr]\frac{1}{\tau_0}+C_3\mathcal{O}(1/k_i) \, ,
\end{align}
where we have ignored higher-order terms in $C_i$. Here, the $1/\tau_0$ terms in $\Delta({\rm Re}[\bar{\mathcal{I}_1}({\bf K^c})])$ and $\Delta({\rm Re}[\bar{\mathcal{I}_3}({\bf K^c})])$ cancel with each other, while the $\mathcal{O}(1/k_i)$ terms do not. As a result, we have the following non-vanishing contributions to the time-ordered variance in Eq. \eqref{time_ordered_graviton_med} that arise when $C_3\neq0$:
\begin{align}
\Delta{\rm Tvar}^{\rm CS}_\lambda({\bf K}^c)&:= -2 \, \Delta({\rm Re}[{\mathcal{I}_1}({\bf K^c})+{\mathcal{I}_2}({\bf K^c})-{\mathcal{I}_3}({\bf K^c})])\notag\\
&=-\frac{C_3 H^{10}}{8 (K_1^c)^3 (K_2^c)^3 (K_3^c)^3 K_4^c \,k^c} \notag\\
&\quad \times \biggl\{\frac{4K_1^c(3K_3^c+K_4^c+k^c)}{(K_3^c+K_4^c+k^c)^3}+\sum_{i=1}^3\biggl[F_i(-K_1^c,K_2^c,K_3^c,K_4^c)+F_i(K_1^c,-K_2^c,K_3^c,K_4^c)\notag\\
&\qquad\qquad +F_i(K_1^c,K_2^c,-K_3^c,K_4^c)+F_i(K_1^c,K_2^c,K_3^c,-K_4^c)-4F_i(K_1^c,K_2^c,K_3^c,K_4^c)\biggr]\notag\\
&\quad +\sum_{j=4}^5\biggl[F_j(K_1^c,K_2^c,-K_3^c,K_4^c)+F_i(K_1^c,K_2^c,K_3^c,-K_4^c)\biggr]\notag\\
&\quad +\sum_{k=6}^7\biggl[F_k(-K_1^c,K_2^c,K_3^c,K_4^c)+F_k(K_1^c,-K_2^c,K_3^c,K_4^c)-2F_k(-K_1^c,-K_2^c,K_3^c,K_4^c)\biggr]\biggr\} \neq 0 \, , 
\end{align}
where
\begin{align}
F_1(x,y,z,w)&:=\frac{2k^c[(x+y)^4-(k^c)^2(x^2+y^2)]}{(x+y+k^c)^2(x+y-k^c)^2(x+y+z+w)^2} \, , \\
F_2(x,y,z,w)&:=\frac{4k^c\{(x+y)^3(xy+yz+zx)-(k^c)^2(xy^2+x^2y+x^2z+y^2z)\}}{(x+y+k^c)^2(x+y-k^c)^2(x+y+z+w)^3} \, , \\
F_3(x,y,z,w)&:=\frac{12k^cxy(x+y)z}{(x+y+k^c)(x+y-k^c)(x+y+z+w)^4} \, ,\\
F_4(x,y,z,w)&:=-2\frac{(x+y)^3(x^2+xy+y^2)-(k^c)^2(x^3+y^3)}{(x+y+k^c)^2(x+y-k^c)^2(k^c+z+w)^2} \, ,\\
F_5(x,y,z,w)&:=-4\frac{[(x+y)^3(x^2+xy+y^2)-(k^c)^2(x^3+y^3)]z}{(x+y+k^c)^2(x+y-k^c)^2(k^c+z+w)^3} \, ,\\
F_6(x,y,z,w)&:=\frac{2k^cy(k^c+y)(k^c+3z+w)}{(x+y+k^c)^2(z+w+k^c)^3} \, ,\\
F_7(x,y,z,w)&:=\frac{2[y^2+(k^c)^2](3z+w+k^c)}{(x+y+k^c)(z+w+k^c)^3} \, .
\end{align}

Similarly to the non-Bunch-Davies scalar modes, one also gets a non-vanishing trispectrum from non-Bunch-Davies tensor modes. To clarify this, let us deform the mode function of the tensor mode as
\begin{align}
\delta\phi^{\rm tensor}_k=\alpha_k\delta\phi^{\rm tensor}_{k,BD}+\beta_k\delta\phi^{\rm tensor *}_{k,BD} \, ,
\end{align}
where $\delta\phi^{\rm tensor}_{k,BD}$ is the Bunch-Davies tensor mode-function in Eq. \eqref{eq:mode_function}, and this time
\begin{align} \label{eq:bolig_tensor}
\tilde\alpha_k&:=1+iD_1 \, , \nonumber\\
\tilde\beta_k&:=D_2+iD_3 \, ,
\end{align}
where for the same reason as above we assume that $D_i$ are real numbers satisfying $|D_i|\ll1$. In this case, we have
\begin{align}
\Delta({\rm Re}[\bar{\mathcal{I}_1}({\bf K^c})])&=D_2\biggl[\frac{(3K_3+K_4+k_c)}{(K_3+K_4+k_c)^3}-\frac{(3K_3+K_4-k_c)}{(K_3+K_4-k_c)^3}\biggr]\frac{1}{\tau_0}+D_3\mathcal{O}(1/k_i) \, , \notag\\
\Delta({\rm Re}[\bar{\mathcal{I}_2}({\bf K^c})])&=D_3\mathcal{O}(1/k_i)\notag \, , \\
\Delta({\rm Re}[\bar{\mathcal{I}_3}({\bf K^c})])&=D_2\biggl[\frac{(3K_3+K_4+k_c)}{(K_3+K_4+k_c)^3}-\frac{(3K_3+K_4-k_c)}{(K_3+K_4-k_c)^3}\biggr]\frac{1}{\tau_0}+D_3\mathcal{O}(1/k_i) \, ,
\end{align}
where we have ignored higher order terms in $D_i$. Here, similarly to the case of non-Bunch-Davies scalar modes, the $1/\tau_0$ terms cancel with each other, and as a result, we obtain the following non-vanishing contributions from $D_3$:
\begin{align}
\Delta{\rm Tvar}^{\rm CS}_\lambda({\bf K}^c)&:= -2 \, \Delta({\rm Re}[{\mathcal{I}_1}({\bf K^c})+{\mathcal{I}_2}({\bf K^c})-{\mathcal{I}_3}({\bf K^c})])\notag\\
&=\frac{D_3 H^{10}}{(K_1^c)^3 (K_2^c)^3 (K_3^c)^3 K_4^c \,k^c} \, (K_1^c+K_2^c)k^c\{(K_1^c+K_2^c)^2[(K_1^c)^2+K_1^cK_2^c+(K_2^c)^2]\notag\\
&\qquad\qquad\ -[(K_1^c)^2-K_1^cK_2^c+(K_2^c)^2](k^c)^2\} \, [(K_3^c+K_4^c)^2(4K_3^c+K_4^c)-K_4^c(k^c)^2]\notag\\
&\quad\ \times\frac{1}{(K_1^c+K_2^c+k^c)^2(K_1^c+K_2^c-k^c)^2(K_3^c+K_4^c+k^c)^3(K_3^c+K_4^c-k^c)^3} \neq 0\, .
\end{align}
We thus find explicitly that imposing the non-Bunch-Davies conditions for the tensor modes also leads to a non-vanishing parity-odd trispectrum.

We then study the impact of the non-Bunch-Davies scalar modes on the trispectrum from the contact diagram. By employing the scalar mode functions in Eq. \eqref{eq:NBD_scalar}, we  get the following modification to the time integral of Eq. \eqref{eq:time_integral_contact}:
\begin{align}
I_{(contact,NBD)}&=2\,{\rm Im}\biggl\{\int_{-\infty}^{\tau_0} d\tau \tilde I_{(contact,BD)}+i C_3\int_{-\infty}^{\tau_0} d\tau \tilde I_{(contact,BD)}\biggl[ e^{-2ik_1\tau}+e^{-2ik_3\tau}\notag\\
&\quad\quad\quad\quad\ +e^{-2ik_2\tau}\frac{1+ik_2\tau}{1-ik_2\tau}+e^{-2ik_4\tau}\frac{1+ik_4\tau}{1-ik_4\tau} - \sum_{i=1}^4 e^{2i  k_i \tau_0} \frac{1-ik_i\tau_0}{1+ik_i\tau_0}\biggr] +\cdots\biggr\}  \, , \notag\\
&\qquad\qquad\qquad\qquad\qquad\qquad\qquad\qquad\qquad\qquad\qquad\qquad 
\end{align}
with
\begin{align}
\tilde I_{(contact,BD)}=\prod_i (1+i k_i \tau_0) \, e^{-i k_i \tau_0} \tau(1-i k_2\tau)(1-i k_4\tau) \, e^{i k_t\tau} \, ,
\end{align}
where $k_t = \sum_{i=1}^4 k_i$. Here, $(\cdots)$ denotes terms of higher order in $C_i$. By dropping those terms and performing the above in-in time integrals taking the limit $\tau_0\rightarrow 0$, we obtain
\begin{align}
I_{(contact,NBD)}&=-2C_3\biggl[S(-k_1,k_2,k_3,k_4)+S(k_1,-k_2,k_3,k_4)+S(k_1,k_2,-k_3,k_4)\notag\\
&\quad\ +S(k_1,k_2,k_3,-k_4)-4S(k_1,k_2,k_3,k_4)\biggr] \neq 0 \, ,
\end{align}
where
\begin{align}
S(k_1,k_2,k_3,k_4):=\frac{1}{k_t^4}(k_1^2+4k_1k_2+3k_2^2+2k_1k_3+4k_2k_3+k_3^2+4k_1k_4+12k_2k_4+4k_3k_4+3k_4^2) \, .
\end{align}
We have thus demonstrated that the simple setup of non-Bunch-Davies initial states in Eq.~\eqref{eq:bolig} (or Eq. \eqref{eq:bolig_tensor})  yields a nonvanishing parity-odd trispectrum, coming both from the graviton-mediated and the contact diagrams. What causes this non-trivial signature is the imaginary part of the coefficient of the negative frequency mode in either the scalar and tensor mode-functions. 

Here, a few final comments are in order. First, we have assumed $C_i, D_i\ll1$, and hence the amplitude of the trispectrum obtained above will be highly suppressed by both $H/M_{CS}\ll1$ and $C_3, D_3\ll1$. Also, as it has been studied in the context of the bispectrum, non-Bunch-Davies initial conditions yield enhancements depending on an explicit momentum-triangle limit (e.g. the folded limit (see e.g.,~\cite{Chen:2006nt,Holman:2007na})) or a particular choice of Bogoliubov coefficients (e.g. that for the so-called alpha vacuum \cite{Christodoulidis:2024ric}). Therefore, it would be interesting to explore the possibility of enhancing the parity-odd trispectrum in specific non-Bunch-Davies setups which, as clarified above, is beyond the scope of our paper.

\section{Summary}
\label{sec:7}

In this paper, we have investigated parity-violating effects in the scalar trispectrum within the framework of dynamical Chern-Simons gravity. Building upon prior studies that considered only GR-based graviton-mediated contributions, we computed the trispectrum arising from both graviton exchange and contact diagrams where at least one interaction vertex is sourced from the gravitational Chern-Simons term. By deriving the relevant cubic and quartic interactions in the spatially flat gauge and using GR-mode graviton propagators, we confirmed that all parity-odd contributions to the scalar trispectrum vanish under the assumption of Bunch-Davies initial conditions.

This result is in alignment with the known no-go theorem forbidding such signatures in the absence of net (logarithmic) infra-red divergences in the in-in time integrals.  In particular, the time scaling of the latters is given by the explicit computation of the interaction vertices (Eqs. \eqref{eq:sst_CS_new} and \eqref{eq:2orderCS}, \eqref{eq:newGR_odd}). These results reinforce the robustness of the no-go theorem in constraining parity-violating signatures in the scalar sector under conventional inflationary assumptions. 

However, by relaxing the Bunch-Davies initial state and introducing a minimal Bogoliubov-type excitation in the scalar and tensor mode-functions, we have demonstrated that a non-zero parity-odd scalar trispectrum can indeed be generated. This non-vanishing signal arises from the interference between positive and negative frequency modes, and its amplitude is linearly dependent on both the Bogoliubov parameter and the ratio $H/M_{CS}$, which remains observationally constrained to be small. While this suppression presents challenges for detectability, our consideration opens a pathway for probing parity-violating physics and non-standard initial states through higher-order correlation functions.

These findings suggest future directions in which the phenomenological consequences of parity violation could be explored more broadly. In particular, studying enhanced configurations of the trispectrum in specific momentum limits, or considering alternative non-Bunch-Davies initial states such as those inspired by $\alpha$-vacua or excited states in effective field theory frameworks, may yield larger observable effects. Additionally, extending the analysis to mixed correlators involving tensor modes could provide complementary observables where parity-odd signals may be amplified. Such avenues remain promising for future investigation.

\section*{Acknowledgements}
We thank David Stefanyszyn, Xi Tong and Yuhang Zhu for discussions on the topics of this research. This work was supported by the grant No.~UMO-2021/42/E/ST9/00260 from the National Science Centre, Poland.

\appendix

\bibliographystyle{hunsrt}
\bibliography{CSV}

@article{Stefanyszyn:2023qov,
    author = "Stefanyszyn, David and Tong, Xi and Zhu, Yuhang",
    title = "{Cosmological correlators through the looking glass: reality, parity, and factorisation}",
    eprint = "2309.07769",
    archivePrefix = "arXiv",
    primaryClass = "hep-th",
    doi = "10.1007/JHEP05(2024)196",
    journal = "JHEP",
    volume = "05",
    pages = "196",
    year = "2024"
}

@article{Stefanyszyn:2025yhq,
    author = "Stefanyszyn, David and Tong, Xi and Zhu, Yuhang",
    title = "{A Match Made in Heaven: Linking Observables in Inflationary Cosmology}",
    eprint = "2505.16071",
    archivePrefix = "arXiv",
    primaryClass = "hep-th",
    month = "5",
    year = "2025"
}

@article{Orlando:2025pkb,
    author = "Orlando, Giorgio",
    title = "{Chern-Simons gravitational term coupled to an isocurvature field}",
    eprint = "2501.09630",
    archivePrefix = "arXiv",
    primaryClass = "astro-ph.CO",
    month = "1",
    year = "2025"
}

@article{Shiraishi:2016mok,
    author = "Shiraishi, Maresuke",
    title = "{Parity violation in the CMB trispectrum from the scalar sector}",
    eprint = "1608.00368",
    archivePrefix = "arXiv",
    primaryClass = "astro-ph.CO",
    reportNumber = "IPMU16-0111",
    doi = "10.1103/PhysRevD.94.083503",
    journal = "Phys. Rev. D",
    volume = "94",
    number = "8",
    pages = "083503",
    year = "2016"
}

@article{Philcox:2023ypl,
    author = "Philcox, Oliver H. E. and Shiraishi, Maresuke",
    title = "{Testing parity symmetry with the polarized cosmic microwave background}",
    eprint = "2308.03831",
    archivePrefix = "arXiv",
    primaryClass = "astro-ph.CO",
    doi = "10.1103/PhysRevD.109.083514",
    journal = "Phys. Rev. D",
    volume = "109",
    number = "8",
    pages = "083514",
    year = "2024"
}

@article{Philcox:2022hkh,
    author = "Philcox, Oliver H. E.",
    title = "{Probing parity violation with the four-point correlation function of BOSS galaxies}",
    eprint = "2206.04227",
    archivePrefix = "arXiv",
    primaryClass = "astro-ph.CO",
    doi = "10.1103/PhysRevD.106.063501",
    journal = "Phys. Rev. D",
    volume = "106",
    number = "6",
    pages = "063501",
    year = "2022"
}

@article{Hou:2022wfj,
    author = "Hou, Jiamin and Slepian, Zachary and Cahn, Robert N.",
    title = "{Measurement of parity-odd modes in the large-scale 4-point correlation function of Sloan Digital Sky Survey Baryon Oscillation Spectroscopic Survey twelfth data release CMASS and LOWZ galaxies}",
    eprint = "2206.03625",
    archivePrefix = "arXiv",
    primaryClass = "astro-ph.CO",
    doi = "10.1093/mnras/stad1062",
    journal = "Mon. Not. Roy. Astron. Soc.",
    volume = "522",
    number = "4",
    pages = "5701--5739",
    year = "2023"
}

@article{Liu:2019fag,
    author = "Liu, Tao and Tong, Xi and Wang, Yi and Xianyu, Zhong-Zhi",
    title = "{Probing P and CP Violations on the Cosmological Collider}",
    eprint = "1909.01819",
    archivePrefix = "arXiv",
    primaryClass = "hep-ph",
    doi = "10.1007/JHEP04(2020)189",
    journal = "JHEP",
    volume = "04",
    pages = "189",
    year = "2020"
}

@article{Thavanesan:2025kyc,
    author = "Thavanesan, Ayngaran",
    title = "{No-go Theorem for Cosmological Parity Violation}",
    eprint = "2501.06383",
    archivePrefix = "arXiv",
    primaryClass = "hep-th",
    month = "1",
    year = "2025"
}

@article{Christodoulidis:2024ric,
    author = "Christodoulidis, Perseas and Gong, Jinn-Ouk and Lin, Wei-Chen and Mylova, Maria and Sasaki, Misao",
    title = "{New shape for cross-bispectra in Chern-Simons gravity}",
    eprint = "2409.09935",
    archivePrefix = "arXiv",
    primaryClass = "hep-th",
    doi = "10.1088/1475-7516/2025/01/037",
    journal = "JCAP",
    volume = "01",
    pages = "037",
    year = "2025"
}

@article{Dyda:2012rj,
    author = "Dyda, Sergei and Flanagan, Eanna E. and Kamionkowski, Marc",
    title = "{Vacuum Instability in Chern-Simons Gravity}",
    eprint = "1208.4871",
    archivePrefix = "arXiv",
    primaryClass = "gr-qc",
    doi = "10.1103/PhysRevD.86.124031",
    journal = "Phys. Rev. D",
    volume = "86",
    pages = "124031",
    year = "2012"
}

@article{Satoh:2010ep,
    author = "Satoh, Masaki",
    title = "{Slow-roll Inflation with the Gauss-Bonnet and Chern-Simons Corrections}",
    eprint = "1008.2724",
    archivePrefix = "arXiv",
    primaryClass = "astro-ph.CO",
    doi = "10.1088/1475-7516/2010/11/024",
    journal = "JCAP",
    volume = "11",
    pages = "024",
    year = "2010"
}

@article{Alexander:2004wk,
    author = "Alexander, Stephon and Martin, Jerome",
    title = "{Birefringent gravitational waves and the consistency check of inflation}",
    eprint = "hep-th/0410230",
    archivePrefix = "arXiv",
    reportNumber = "SLAC-PUB-10816",
    doi = "10.1103/PhysRevD.71.063526",
    journal = "Phys. Rev. D",
    volume = "71",
    pages = "063526",
    year = "2005"
}

@article{Lue:1998mq,
    author = "Lue, Arthur and Wang, Li-Min and Kamionkowski, Marc",
    title = "{Cosmological signature of new parity violating interactions}",
    eprint = "astro-ph/9812088",
    archivePrefix = "arXiv",
    reportNumber = "CU-TP-926, CAL-675",
    doi = "10.1103/PhysRevLett.83.1506",
    journal = "Phys. Rev. Lett.",
    volume = "83",
    pages = "1506--1509",
    year = "1999"
}

@article{Achucarro:2022qrl,
    author = "Ach\'ucarro, Ana and others",
    title = "{Inflation: Theory and Observations}",
    eprint = "2203.08128",
    archivePrefix = "arXiv",
    primaryClass = "astro-ph.CO",
    month = "3",
    year = "2022"
}

@article{Cabass:2022rhr,
    author = "Cabass, Giovanni and Jazayeri, Sadra and Pajer, Enrico and Stefanyszyn, David",
    title = "{Parity violation in the scalar trispectrum: no-go theorems and yes-go examples}",
    eprint = "2210.02907",
    archivePrefix = "arXiv",
    primaryClass = "hep-th",
    doi = "10.1007/JHEP02(2023)021",
    journal = "JHEP",
    volume = "02",
    pages = "021",
    year = "2023"
}

@article{Tian:2015vda,
    author = "Tian, David Wenjie and Booth, Ivan",
    title = "{Lovelock\textendash{}Brans\textendash{}Dicke gravity}",
    eprint = "1502.05695",
    archivePrefix = "arXiv",
    primaryClass = "gr-qc",
    doi = "10.1088/0264-9381/33/4/045001",
    journal = "Class. Quant. Grav.",
    volume = "33",
    number = "4",
    pages = "045001",
    year = "2016"
}

@article{Seery:2006vu,
    author = "Seery, David and Lidsey, James E. and Sloth, Martin S.",
    title = "{The inflationary trispectrum}",
    eprint = "astro-ph/0610210",
    archivePrefix = "arXiv",
    doi = "10.1088/1475-7516/2007/01/027",
    journal = "JCAP",
    volume = "01",
    pages = "027",
    year = "2007"
}

@article{Chen:2017ryl,
    author = "Chen, Xingang and Wang, Yi and Xianyu, Zhong-Zhi",
    title = "{Schwinger-Keldysh Diagrammatics for Primordial Perturbations}",
    eprint = "1703.10166",
    archivePrefix = "arXiv",
    primaryClass = "hep-th",
    doi = "10.1088/1475-7516/2017/12/006",
    journal = "JCAP",
    volume = "12",
    pages = "006",
    year = "2017"
}

@article{Maldacena:2002vr,
    author = "Maldacena, Juan Martin",
    title = "{Non-Gaussian features of primordial fluctuations in single field inflationary models}",
    eprint = "astro-ph/0210603",
    archivePrefix = "arXiv",
    doi = "10.1088/1126-6708/2003/05/013",
    journal = "JHEP",
    volume = "05",
    pages = "013",
    year = "2003"
}

@article{Creque-Sarbinowski:2023wmb,
    author = "Creque-Sarbinowski, Cyril and Alexander, Stephon and Kamionkowski, Marc and Philcox, Oliver",
    title = "{Parity-violating trispectrum from Chern-Simons gravity}",
    eprint = "2303.04815",
    archivePrefix = "arXiv",
    primaryClass = "astro-ph.CO",
    doi = "10.1088/1475-7516/2023/11/029",
    journal = "JCAP",
    volume = "11",
    pages = "029",
    year = "2023"
}

@article{Holman:2007na,
    author = "Holman, R. and Tolley, Andrew J.",
    title = "{Enhanced Non-Gaussianity from Excited Initial States}",
    eprint = "0710.1302",
    archivePrefix = "arXiv",
    primaryClass = "hep-th",
    reportNumber = "PI-COSMO-64",
    doi = "10.1088/1475-7516/2008/05/001",
    journal = "JCAP",
    volume = "05",
    pages = "001",
    year = "2008"
}

@article{Jackiw:2003pm,
    author = "Jackiw, R. and Pi, S. Y.",
    title = "{Chern-Simons modification of general relativity}",
    eprint = "gr-qc/0308071",
    archivePrefix = "arXiv",
    reportNumber = "MIT-CTP-3409, BUHEP-03-18",
    doi = "10.1103/PhysRevD.68.104012",
    journal = "Phys. Rev. D",
    volume = "68",
    pages = "104012",
    year = "2003"
}

@article{Bartolo:2017szm,
    author = "Bartolo, Nicola and Orlando, Giorgio",
    title = "{Parity breaking signatures from a Chern-Simons coupling during inflation: the case of non-Gaussian gravitational waves}",
    eprint = "1706.04627",
    archivePrefix = "arXiv",
    primaryClass = "astro-ph.CO",
    doi = "10.1088/1475-7516/2017/07/034",
    journal = "JCAP",
    volume = "07",
    pages = "034",
    year = "2017"
}

@article{Gong:2023kpe,
    author = "Gong, Jinn-Ouk and Mylova, Maria and Sasaki, Misao",
    title = "{New shape of parity-violating graviton non-Gaussianity}",
    eprint = "2303.05178",
    archivePrefix = "arXiv",
    primaryClass = "hep-th",
    reportNumber = "APCTP-Pre2023-003, YITP-23-23",
    doi = "10.1007/JHEP10(2023)140",
    journal = "JHEP",
    volume = "10",
    pages = "140",
    year = "2023"
}

@article{Chen:2006nt,
    author = "Chen, Xingang and Huang, Min-xin and Kachru, Shamit and Shiu, Gary",
    title = "{Observational signatures and non-Gaussianities of general single field inflation}",
    eprint = "hep-th/0605045",
    archivePrefix = "arXiv",
    reportNumber = "SLAC-PUB-11840, MAD-TH-06-3, UFIFT-HEP-06-9, SU-ITP-06-12, CU-TP-1147",
    doi = "10.1088/1475-7516/2007/01/002",
    journal = "JCAP",
    volume = "01",
    pages = "002",
    year = "2007"
}

@article{Tanaka:2000jw,
    author = "Tanaka, Takahiro",
    title = "{A Comment on transPlanckian physics in inflationary universe}",
    eprint = "astro-ph/0012431",
    archivePrefix = "arXiv",
    month = "12",
    year = "2000"
}

\end{document}